\documentclass[prd,showkeys,floatfix,twocolumn,amsmath,amssymb,floatfix]{revtex4}
\usepackage{graphicx}
\usepackage{multirow}
\usepackage{subfigure}
\usepackage[colorlinks,citecolor=blue,anchorcolor=red,menucolor=red,linkcolor=red,filecolor=red,runcolor=red,urlcolor=blue,frenchlinks=red]{hyperref}
\usepackage{enumitem}

\makeatletter
\renewcommand{\@thesubfigure}{\hskip\subfiglabelskip}
\makeatother
\allowdisplaybreaks[3]

\begin{document}
\title{Light double-gluon hybrid states from QCD sum rules}
%

\author{Niu Su$^1$}
\author{Hua-Xing Chen$^1$}
\email{hxchen@seu.edu.cn}
\author{Wei Chen$^2$}
\email{chenwei29@mail.sysu.edu.cn}
\author{Shi-Lin Zhu$^3$}
\email{zhusl@pku.edu.cn}

\affiliation{
$^1$School of Physics, Southeast University, Nanjing 210094, China\\
$^2$School of Physics, Sun Yat-Sen University, Guangzhou 510275, China\\
$^3$School of Physics and Center of High Energy Physics, Peking University, Beijing 100871, China}
\begin{abstract}
We study the double-gluon hybrid states with the quark-gluon contents $\bar q q gg$ ($q=u/d$) and $\bar s s gg$. We construct twelve double-gluon hybrid currents with various quantum numbers, five of which are found to be zero due to some internal symmetries between the two gluon fields. We use the rest seven currents to perform QCD sum rule analyses. Especially, the masses of the double-gluon hybrid states with the exotic quantum number $J^{PC} = 2^{+-}$ are calculated to be $M_{|\bar q q gg;2^{+-}\rangle} = 2.26^{+0.20}_{-0.25}$~GeV and $M_{|\bar s s gg;2^{+-}\rangle} = 2.38^{+0.19}_{-0.25}$~GeV. Their two- and three-meson decay patterns are also investigated.
\end{abstract}
\maketitle
\pagenumbering{arabic}

\section{Introduction}
\label{sec:intro}

A hybrid state is composed of one valence quark, one valence antiquark, and one or more valence gluons. Some of the hybrid states have the exotic quantum numbers $J^{PC} =0^{--}/0^{+-}/1^{-+}/2^{+-}/3^{-+}/4^{+-}/\cdots$ that can not be accessed by the conventional $\bar q q$ mesons~\cite{pdg}. These hybrid states are of particular interests, which have been studied in various experimental and theoretical investigations in the past half century~\cite{Chen:2022asf,Klempt:2007cp,Meyer:2015eta,Amsler:2004ps,Bugg:2004xu,Meyer:2010ku,Briceno:2017max,COMPASS:2018uzl,JPAC:2018zyd,Ketzer:2019wmd,Jin:2021vct,Meng:2022ozq}.

Up to now there are four candidates observed in experiments with the exotic quantum number $J^{PC} = 1^{-+}$, including the $\pi_1(1400)$~\cite{IHEP-Brussels-LosAlamos-AnnecyLAPP:1988iqi}, $\pi_1(1600)$~\cite{E852:1998mbq,COMPASS:2009xrl}, and $\pi_1(2015)$~\cite{E852:2004gpn} of $I^GJ^{PC} = 1^-1^{-+}$ as well as the $\eta_1(1855)$~\cite{BESIII:2022riz,BESIII:2022qzu} of $I^GJ^{PC} = 0^+1^{-+}$. Especially, the last one was recently reported by the BESIII collaboration in the $\eta \eta^\prime$ invariant mass spectrum of the $J/\psi \to \gamma \eta \eta^\prime$ decay with a statistical significance larger than $19\sigma$~\cite{BESIII:2022riz,BESIII:2022qzu}. These candidates are possible single-gluon hybrid states, which are composed of one valence quark-antiquark pair and one valence gluon. Besides, they may also be explained as the compact tetraquark states and the hadronic molecular states~\cite{Chen:2008qw,Chen:2008ne,Zhang:2019ykd,Dong:2022cuw,Yang:2022lwq,Wan:2022xkx,Wang:2022sib,Su:2022eun,Yu:2022wtu}, etc.

The above candidates have been intensively studied within the hybrid picture using various theoretical methods and models, such as the MIT bag model~\cite{Barnes:1977hg,Hasenfratz:1980jv,Chanowitz:1982qj}, flux-tube model~\cite{Isgur:1983wj,Close:1994hc,Page:1998gz,Burns:2006wz,Qiu:2022ktc}, constituent gluon model~\cite{Szczepaniak:2001rg,Iddir:2007dq,Guo:2007sm}, AdS/QCD model~\cite{Andreev:2012hw,Bellantuono:2014lra}, Dyson-Schwinger equation~\cite{Xu:2018cor}, lattice QCD~\cite{Michael:1985ne,McNeile:1998cp,Juge:2002br,Lacock:1996ny,MILC:1997usn,Bernard:2003jd,Hedditch:2005zf,Dudek:2009qf,Dudek:2010wm,Dudek:2013yja,Chen:2022isv}, and QCD sum rules~\cite{Balitsky:1982ps,Govaerts:1983ka,Kisslinger:1995yw,Chetyrkin:2000tj,Jin:2002rw,Narison:2009vj,Huang:2010dc,Chen:2010ic,Huang:2016upt,Li:2021fwk}, etc~\cite{Shastry:2022mhk}. However, their nature is still elusive due to our poor understanding of the ``valence'' gluon:
\begin{itemize}

\item It is not easy to experimentally identify the hybrid states unambiguously, and there is currently no definite experimental evidence on their existence. This is partly because that it is rather difficult to differentiate the hybrid states from the compact tetraquark states and the hadronic molecular states. This tough problem needs to be solved by experimentalists and theorists together in future.

\item It is also not easy to theoretically define the gluon degree of freedom. There have been some proposals to construct glueballs and hybrid states using the constituent gluons~\cite{Horn:1977rq,Coyne:1980zd,Chanowitz:1980gu,Barnes:1981ac,Cornwall:1982zn,Cho:2015rsa}, but a precise definition of the constituent gluon is still lacking.
\end{itemize}

In this paper we shall further investigate the double-gluon hybrid states, which are composed of one valence quark-antiquark pair and two valence gluons. Some previous QCD sum rule studies on these states can be found in Refs.~\cite{Chen:2021smz,Tang:2021zti}. In this paper we shall study the double-gluon hybrid states with the quark-gluon contents $\bar q q gg$ ($q=u/d$) and $\bar s s gg$. We shall construct twelve double-gluon hybrid currents with various quantum numbers, five of which are found to be zero due to some internal symmetries between the two gluon fields. We shall use the rest seven currents to perform QCD sum rule analyses. Especially, the masses of the double-gluon hybrid states with the exotic quantum number $J^{PC} = 2^{+-}$ are calculated to be smaller than 3.0~GeV:
\begin{eqnarray}
\nonumber M_{|\bar q q gg;1^+2^{+-}\rangle} = M_{|\bar q q gg;0^-2^{+-}\rangle} &=& 2.26^{+0.20}_{-0.25}{\rm~GeV} \, ,
\\ \nonumber M_{|\bar s s gg;0^-2^{+-}\rangle} &=& 2.38^{+0.19}_{-0.25}{\rm~GeV} \, .
\end{eqnarray}
These mass values are accessible in the BESIII, GlueX, LHC, and PANDA experiments. In this paper we shall also investigate their possible decay patterns for both the two- and three-meson final states.

This paper is organized as follows. In Sec.~\ref{sec:current} we construct twelve double-gluon hybrid currents, five of which are found to be zero. We use the rest seven currents to perform QCD sum rule analyses in Sec.~\ref{sec:sumrule}, and then perform numerical analyses in Sec.~\ref{sec:numerical}. The obtained results are summarized and discussed in Sec.~\ref{sec:summary}.

\section{Double-gluon hybrid currents}
\label{sec:current}

In this section we construct the double-gluon hybrid currents by using the gluon field strength tensor $G^n_{\mu\nu}(x)$, the dual gluon field strength tensor $\tilde G^n_{\mu\nu} = G^{n,\rho\sigma} \times \epsilon_{\mu\nu\rho\sigma}/2$, the light quark field $q_a(x)$, and the light antiquark field $\bar q_a(x)$. In these expressions, $a=1\cdots3$ and $n=1\cdots8$ are color indices, and $\mu,\nu,\rho,\sigma$ are Lorentz indices.

Some of the double-gluon hybrid currents have been constructed in Ref.~\cite{Chen:2021smz}:
\begin{eqnarray}
J_{0^{++}} &=& \bar q_a \gamma_5 \lambda_n^{ab} q_b~d^{npq}~g_s^2 G_p^{\mu\nu} \tilde G_{q,\mu\nu} \, ,
\label{def:0pp}
\\
J_{0^{+-}} &=& \bar q_a \gamma_5 \lambda_n^{ab} q_b~f^{npq}~g_s^2 G_p^{\mu\nu} \tilde G_{q,\mu\nu} \, ,
\label{def:0pn}
\\
J_{0^{-+}} &=& \bar q_a \gamma_5 \lambda_n^{ab} q_b~d^{npq}~g_s^2 G_p^{\mu\nu} G_{q,\mu\nu} \, ,
\label{def:0np}
\\
J_{0^{--}} &=& \bar q_a \gamma_5 \lambda_n^{ab} q_b~f^{npq}~g_s^2 G_p^{\mu\nu} G_{q,\mu\nu} \, ,
\label{def:0nn}
\\ \nonumber
J^{\alpha\beta}_{1^{++}} &=& \bar q_a \gamma_5 \lambda_n^{ab} q_b~d^{npq}~g_s^2 G_p^{\alpha\mu} \tilde G_{q,\mu}^\beta - \{ \alpha \leftrightarrow \beta \} \, ,
\\
\label{def:1pp}
\\ \nonumber
J^{\alpha\beta}_{1^{+-}} &=& \bar q_a \gamma_5 \lambda_n^{ab} q_b~f^{npq}~g_s^2 G_p^{\alpha\mu} \tilde G_{q,\mu}^\beta - \{ \alpha \leftrightarrow \beta \} \, ,
\\
\label{def:1pn}
\\ \nonumber
J^{\alpha\beta}_{1^{-+}} &=& \bar q_a \gamma_5 \lambda_n^{ab} q_b~d^{npq}~g_s^2 G_p^{\alpha\mu} G_{q,\mu}^\beta - \{ \alpha \leftrightarrow \beta \} \, ,
\\
\label{def:1np}
\\ \nonumber
J^{\alpha\beta}_{1^{--}} &=& \bar q_a \gamma_5 \lambda_n^{ab} q_b~f^{npq}~g_s^2 G_p^{\alpha\mu} G_{q,\mu}^\beta - \{ \alpha \leftrightarrow \beta \} \, ,
\\
\label{def:1nn}
\\
J^{\alpha_1\beta_1,\alpha_2\beta_2}_{2^{++}} &=& \bar q_a \gamma_5 \lambda_n^{ab} q_b~d^{npq}~\mathcal{S}[ g_s^2 G_p^{\alpha_1\beta_1} \tilde G_q^{\alpha_2\beta_2} ] \, ,
\label{def:2pp}
\\
J^{\alpha_1\beta_1,\alpha_2\beta_2}_{2^{+-}} &=& \bar q_a \gamma_5 \lambda_n^{ab} q_b~f^{npq}~\mathcal{S}[ g_s^2 G_p^{\alpha_1\beta_1} \tilde G_q^{\alpha_2\beta_2} ] \, ,
\label{def:2pn}
\\
J^{\alpha_1\beta_1,\alpha_2\beta_2}_{2^{-+}} &=& \bar q_a \gamma_5 \lambda_n^{ab} q_b~d^{npq}~\mathcal{S}[ g_s^2 G_p^{\alpha_1\beta_1} G_q^{\alpha_2\beta_2} ] \, ,
\label{def:2np}
\\
J^{\alpha_1\beta_1,\alpha_2\beta_2}_{2^{--}} &=& \bar q_a \gamma_5 \lambda_n^{ab} q_b~f^{npq}~\mathcal{S}[ g_s^2 G_p^{\alpha_1\beta_1} G_q^{\alpha_2\beta_2} ] \, ,
\label{def:2nn}
\end{eqnarray}
where $\mathcal{S}$ represents symmetrization and subtracting trace terms in the two sets $\{\alpha_1 \alpha_2\}$ and $\{\beta_1 \beta_2\}$ simultaneously.

The above double-gluon hybrid currents all contain the color-octet quark-antiquark field $\bar q_a \gamma_5 \lambda_n^{ab} q_b$ with the $S$-wave spin-parity quantum number $J^P = 0^-$, so these currents may couple to the lowest-lying double-gluon hybrid states. Their color structure is
\begin{eqnarray}
\mathbf{3}_q \otimes \mathbf{\bar 3}_{\bar q} \otimes \mathbf{8}_g \otimes \mathbf{8}_g &\rightarrow& \mathbf{8}_{\bar q q} \otimes \mathbf{8}_g \otimes \mathbf{8}_g
\\ \nonumber &\rightarrow& \mathbf{1}^S_{\bar q q gg} \oplus \mathbf{1}^A_{\bar q q gg} \, ,
\end{eqnarray}
where $\mathbf{1}^S_{\bar q q gg}$ denotes the symmetric color configuration $d^{npq}\bar q_a \lambda_n^{ab} q_b G_p G_q$ and $\mathbf{1}^A_{\bar q q gg}$ denotes the antisymmetric color configuration $f^{npq} \bar q_a \lambda_n^{ab} q_b G_p G_q$, with $d^{npq}$ and $f^{npq}$ the totally symmetric and antisymmetric $SU(3)$ structure constants respectively.

More double-gluon hybrid currents can be constructed by combining the color-octet quark-antiquark fields
\begin{equation}
\bar q_a \lambda_n^{ab} q_b \, , \,  \bar q_a \lambda_n^{ab} \gamma_\mu q_b \, , \, \bar q_a \lambda_n^{ab} \gamma_\mu \gamma_5 q_b \, , \, \bar q_a \lambda_n^{ab} \sigma_{\mu\nu} q_b \, ,
\end{equation}
and the color-octet double-gluon fields
\begin{equation}
d^{npq} G_p^{\alpha\beta} G_q^{\gamma\delta} \, , \, f^{npq} G_p^{\alpha\beta} G_q^{\gamma\delta} \, ,
\end{equation}
together with some Lorentz matrices $\Gamma^{\mu\nu\cdots\alpha\beta\gamma\delta}$. Note that some of these currents can mix with the above currents defined in Eqs.~(\ref{def:0pp}-\ref{def:2nn}) when they have the same spin-parity quantum numbers.

Each of the four currents, $J_{0^{\pm\pm}}$ defined in Eqs.~(\ref{def:0pp}-\ref{def:0nn}), couples to either the positive- or negative-parity hybrid state. For example, the current $J_{0^{++}}$ couples to the double-gluon hybrid state of $J^{PC} = 0^{++}$ through
\begin{eqnarray}
\langle 0| J_{0^{++}} | X ; 0^{++} \rangle &=& f_{| X ; 0^{++} \rangle}  \, ,
\label{eq:0pp}
\end{eqnarray}
while it does not couple to the state of $J^{PC} = 0^{-+}$.

The other eight currents of spin-$J$, $J^{\alpha\beta}_{1^{\pm\pm}}$ and $J^{\alpha_1\beta_1,\alpha_2\beta_2}_{2^{\pm\pm}}$ defined in Eqs.~(\ref{def:1pp}-\ref{def:2nn}), have $2 \times J$ Lorentz indices with certain symmetries, {\it e.g.}, the spin-2 current $J^{\alpha_1\beta_1,\alpha_2\beta_2}_{2^{++}}$ has four Lorentz indices, satisfying
\begin{equation}
J^{\alpha_1\beta_1,\alpha_2\beta_2}_{2^{++}} = - J^{\beta_1\alpha_1,\alpha_2\beta_2}_{2^{++}} = - J^{\alpha_1\beta_1,\beta_2\alpha_2}_{2^{++}} = J^{\beta_1\alpha_1,\beta_2\alpha_2}_{2^{++}} \, .
\end{equation}
Each of these eight currents couples to both the positive- and negative-parity hybrid states. We briefly explain how to deal with them as follows.

We use the current $J^{\alpha\beta} = \bar c \sigma^{\alpha\beta} c$ as an example, which can be separated into ($\alpha,\beta=0,1,2,3$ and $i,j=1,2,3$):
\begin{equation}
J^{\alpha\beta} = \bar c \sigma^{\alpha\beta} c
\left\{\begin{array}{l}
\bar c \sigma^{ij} c \, , \, {P=+} \, ,
\\[1mm]
\bar c \sigma^{0i} c \, , \, {P=-} \, .
\end{array}\right.
\end{equation}
Therefore, it couples to both the positive- and negative-parity charmonia through
\begin{eqnarray}
\langle 0 | J^{\alpha\beta} | h_c (\epsilon,p) \rangle &=& i f^T_{h_c} \epsilon^{\alpha\beta\mu\nu} \epsilon_\mu p_\nu \, ,
\\ \langle 0 | J^{\alpha\beta} | J/\psi (\epsilon,p) \rangle &=& i f^T_{J/\psi} (p^\alpha\epsilon^\beta - p^\beta\epsilon^\alpha) \, ,
\end{eqnarray}
with $f^T_{h_c}$ and $f^T_{J/\psi}$ the decay constants. We can further isolate the $h_c$ at the hadron level by investigating the two-point correlation function containing
\begin{eqnarray}
&& \langle 0 | J^{\alpha\beta} | h_c \rangle \langle h_c | \left(J^{\alpha^\prime\beta^\prime}\right)^\dagger | 0 \rangle
\\ \nonumber &=& \left( f^T_{h_c} \right)^2 \epsilon^{\alpha\beta\mu\nu} \epsilon_\mu p_\nu \epsilon^{\alpha^\prime\beta^\prime\mu^\prime\nu^\prime} \epsilon^*_{\mu^\prime} p_{\nu^\prime}
\\ \nonumber &=& - \left( f^T_{h_c} \right)^2 ~ p^2 ~ \left( g^{\alpha \alpha^\prime} g^{\beta \beta^\prime} - g^{\alpha \beta^\prime} g^{\beta \alpha^\prime} \right) + \cdots \, ,
\end{eqnarray}
while the correlation function of $J/\psi$ does not contain this coefficient. Instead, the $J/\psi$ can be isolated more easily through the dual current
\begin{equation}
\tilde J^{\alpha\beta} = \epsilon^{\alpha\beta\gamma\delta} \times J_{\gamma\delta} \, ,
\end{equation}
which couples to the $J/\psi$ and $h_c$ in the opposite ways:
\begin{eqnarray}
\langle 0 | \tilde J^{\alpha\beta} | J/\psi (\epsilon,p) \rangle &=& i \tilde f^T_{J/\psi} \epsilon^{\alpha\beta\mu\nu} \epsilon_\mu p_\nu \, ,
\\ \langle 0 | \tilde J^{\alpha\beta} | h_c (\epsilon,p) \rangle &=& i \tilde f^T_{h_c} (p^\alpha\epsilon^\beta - p^\beta\epsilon^\alpha) \, .
\end{eqnarray}
Accordingly, we can use the two currents $J^{\alpha\beta}$ and $\tilde J^{\alpha\beta}$ to separately investigate the two charmonia $h_c$ and $J/\psi$.

The above process can be applied to generally investigate the currents $J^{\alpha\beta}_{1^{\pm\pm}}$ and $J^{\alpha_1\beta_1,\alpha_2\beta_2}_{2^{\pm\pm}}$, with the couplings defined as:
\begin{eqnarray}
\langle 0 | J_{1^{\pm\pm}}^{\alpha\beta} | X ; 1^{\pm\pm}(\epsilon,p) \rangle &=& i f_{1^{\pm\pm}} \epsilon^{\alpha\beta\mu\nu} \epsilon_\mu p_\nu \, ,
\\
\langle 0 | J^{\alpha_1\beta_1,\alpha_2\beta_2}_{2^{\pm\pm}} | X ; 2^{\pm\pm}(\epsilon,p) \rangle &=& i f_{2^{\pm\pm}} \epsilon_{\mu_1 \mu_2} p_{\nu_1} p_{\nu_2}
\\ \nonumber && \times \mathcal{S}[ \epsilon^{\alpha_1 \beta_1 \mu_1 \nu_1} \epsilon^{\alpha_2 \beta_2 \mu_2 \nu_2} ] \, .
\end{eqnarray}
Here $|X ; 1^{\pm\pm}\rangle$ and $|X ; 2^{\pm\pm}\rangle$ are the double-gluon hybrid states that have the same parities as the components $J^{ij}_{1^{\pm\pm}}$ and $J^{i_1j_1,i_2j_2}_{2^{\pm\pm}}$ ($i,i_1,i_2,j,j_1,j_2=1,2,3$), respectively.

Before performing QCD sum rule analyses, let us further examine the twelve currents $J^{\cdots}_{0/1/2^{\pm\pm}}$ defined in Eqs.~(\ref{def:0pp}-\ref{def:2nn}). We find some internal symmetries between the two gluon fields, which make the five currents $J^{\cdots}_{0^{\pm-}/1^{\pm+}/2^{--}}$ vanish:
\begin{itemize}

\item The current $J_{0^{--}}$ contains two totally antisymmetric gluon fields (their Lorentz indices are symmetric and their color coefficient $f^{npq}$ is antisymmetric), so it vanishes due to the Bose-Einstein statistics:
\begin{eqnarray}
J_{0^{--}} &=& \cdots \times f^{npq} ~ G_p^{\mu\nu} G_{q,\mu\nu}
\\ \nonumber &=& \cdots \times f^{nqp} ~ G_q^{\mu\nu} G_{p,\mu\nu}
\\ \nonumber &=& - \cdots \times f^{npq} ~ G_p^{\mu\nu} G_{q,\mu\nu}
\\ \nonumber &=& - J_{0^{--}} \, .
\end{eqnarray}

\item The current $J_{0^{+-}}$ vanishes through some similar deductions:
\begin{eqnarray}
J_{0^{+-}} &=& \cdots \times f^{npq} ~ G_p^{\mu\nu} \tilde G_{q,\mu\nu}
\\ \nonumber &=& \cdots \times f^{nqp} ~ G_q^{\mu\nu} \tilde G_{p,\mu\nu}
\\ \nonumber &=& - \cdots \times f^{npq} ~ \tilde G_p^{\mu\nu} G_{q,\mu\nu}
\\ \nonumber &=& - \cdots \times f^{npq} ~ G_p^{\mu\nu} \tilde G_{q,\mu\nu}
\\ \nonumber &=& - J_{0^{+-}} \, .
\end{eqnarray}

\item The current $J^{\alpha\beta}_{1^{-+}}$ contains two totally antisymmetric gluon fields (their Lorentz indices are antisymmetric and their color coefficient $d^{npq}$ is symmetric), so it vanishes due to the Bose-Einstein statistics:
\begin{eqnarray}
J^{\alpha\beta}_{1^{-+}} &=& \cdots \times d^{npq}~\left( G_p^{\alpha\mu} G_{q,\mu}^\beta - G_p^{\beta\mu} G_{q,\mu}^\alpha \right)
\\ \nonumber &=& \cdots \times d^{nqp}~\left( G_q^{\alpha\mu} G_{p,\mu}^\beta - G_q^{\beta\mu} G_{p,\mu}^\alpha \right)
\\ \nonumber &=& - \cdots \times d^{npq}~\left( G_p^{\alpha\mu} G_{q,\mu}^\beta - G_p^{\beta\mu} G_{q,\mu}^\alpha \right)
\\ \nonumber &=& - J^{\alpha\beta}_{1^{-+}} \, .
\end{eqnarray}

\item The current $J^{\alpha\beta}_{1^{++}}$ vanishes because:
\begin{eqnarray}
&& {J^{01}_{1^{++}} / (g_s^2 \bar q_a \gamma_5 \lambda_n^{ab} q_b)}
\\ \nonumber &=& d^{npq}~\left( G_p^{0\mu} \tilde G_q^{1\mu} - G_p^{1\mu} \tilde G_q^{0\mu} \right)
\\ \nonumber &=& d^{npq}~\left( G_p^{02} \tilde G_q^{12} + G_p^{03} \tilde G_q^{13} - G_p^{12} \tilde G_q^{02} - G_p^{13} \tilde G_q^{03} \right)
\\ \nonumber &=& {d^{npq} \over2}~\left( G_p^{02} G_q^{03} - G_p^{03} G_q^{02} + G_p^{12} G_q^{13} - G_p^{13} G_q^{12} \right)
\\ \nonumber &=& 0 \, .
\end{eqnarray}

\item The current $J^{\alpha_1\beta_1,\alpha_2\beta_2}_{2^{--}}$ contains two totally antisymmetric gluon fields (their Lorentz indices are symmetric and their color coefficient $f^{npq}$ is antisymmetric), so it vanishes due to the Bose-Einstein statistics:
\begin{eqnarray}
J^{\alpha_1\beta_1,\alpha_2\beta_2}_{2^{--}} &=& \cdots \times f^{npq} ~ \mathcal{S}[ G_p^{\alpha_1\beta_1} G_q^{\alpha_2\beta_2} ]
\\ \nonumber &=& \cdots \times f^{nqp} ~ \mathcal{S}[ G_q^{\alpha_1\beta_1} G_p^{\alpha_2\beta_2} ]
\\ \nonumber &=& - \cdots \times f^{npq} ~ \mathcal{S}[ G_p^{\alpha_2\beta_2} G_q^{\alpha_1\beta_1} ]
\\ \nonumber &=& - \cdots \times f^{npq} ~ \mathcal{S}[ G_p^{\alpha_1\beta_1} G_q^{\alpha_2\beta_2} ]
\\ \nonumber &=& - J^{\alpha_1\beta_1,\alpha_2\beta_2}_{2^{--}} \, .
\end{eqnarray}

\end{itemize}
Interestingly, the current $J^{\alpha_1\beta_1,\alpha_2\beta_2}_{2^{+-}}$ does not vanish, {\it e.g.},
\begin{eqnarray}
J^{01,02}_{2^{+-}} &=& \cdots \times f^{npq} ~ \mathcal{S}[ G_p^{01} \tilde G_q^{02} ]
\\ \nonumber &=& \cdots \times f^{npq} ~\left( G_p^{01} \tilde G_q^{02} + G_p^{02} \tilde G_q^{01} \right)
\\ \nonumber &=& \cdots \times {f^{npq} \over 2} ~\left( - G_p^{01} G_q^{13} + G_p^{02} G_q^{23} \right)
\\ \nonumber &\neq& 0 \, .
\end{eqnarray}

%
\section{QCD sum rule analyses}
\label{sec:sumrule}
%

In this section we use the seven currents $J^{\cdots}_{0^{\pm+}/1^{\pm-}/2^{\pm+}/2^{+-}}$ to perform QCD sum rule analyses. We use the current $J_{0^{++}}$ defined in Eq.~(\ref{def:0pp}) as an example. Based on Eq.~(\ref{eq:0pp}), we study its two-point correlation function
\begin{equation}
\Pi(q^2) = i \int d^4x e^{iqx} \langle 0 | {\bf T}[J_{0^{++}}(x) J_{0^{++}}^\dagger(0)] | 0 \rangle \, ,
\label{eq:correlation}
\end{equation}
at both the hadron and quark-gluon levels.

At the hadron level, Eq.~(\ref{eq:correlation}) can be expressed by the dispersion relation as
%
\begin{equation}
\Pi(q^2) = \int^\infty_{s_<}\frac{\rho(s)}{s-q^2-i\varepsilon}ds \, ,
\label{eq:hadron}
\end{equation}
%
where $s_< = 4 m_q^2$ is the physical threshold. The spectral density $\rho(s) \equiv {\rm Im}\Pi(s)/\pi$ is parameterized as one pole dominance for the possibly-existing ground state $X \equiv | X; 0^{++} \rangle$ together with a continuum contribution
%
\begin{eqnarray}
\nonumber \rho_{\rm phen}(s) &\equiv& \sum_n\delta(s-M^2_n) \langle 0| J_{0^{++}} | n\rangle \langle n| J_{0^{++}}^\dagger |0 \rangle
\\ &=& f^2_X \delta(s-M^2_X) + \rm{continuum} \, .
\label{eq:rho}
\end{eqnarray}
%

At the quark-gluon level, Eq.~(\ref{eq:correlation}) can be calculated by the method of operator product expansion (OPE), from which we can extract the OPE spectral density $\rho_{\rm OPE}(s)$. We perform the Borel transformation at both the hadron and quark-gluon levels, and approximate the continuum using $\rho_{\rm OPE}(s)$ above the threshold value $s_0$, from which we obtain the sum rule equation
%
\begin{equation}
\Pi(s_0, M_B^2) \equiv f^2_X e^{-M_X^2/M_B^2} = \int^{s_0}_{s_<} e^{-s/M_B^2}\rho_{\rm OPE}(s)ds \, .
\label{eq:fin}
\end{equation}
%
It can be used to further derive
%
\begin{equation}
M^2_X(s_0, M_B) = \frac{\int^{s_0}_{s_<} e^{-s/M_B^2}s\rho_{\rm OPE}(s)ds}{\int^{s_0}_{s_<} e^{-s/M_B^2}\rho_{\rm OPE}(s)ds} \, .
\label{eq:LSR}
\end{equation}
%

\begin{figure}[hbtp]
\begin{center}
\subfigure[(a)]{
\scalebox{0.12}{\includegraphics{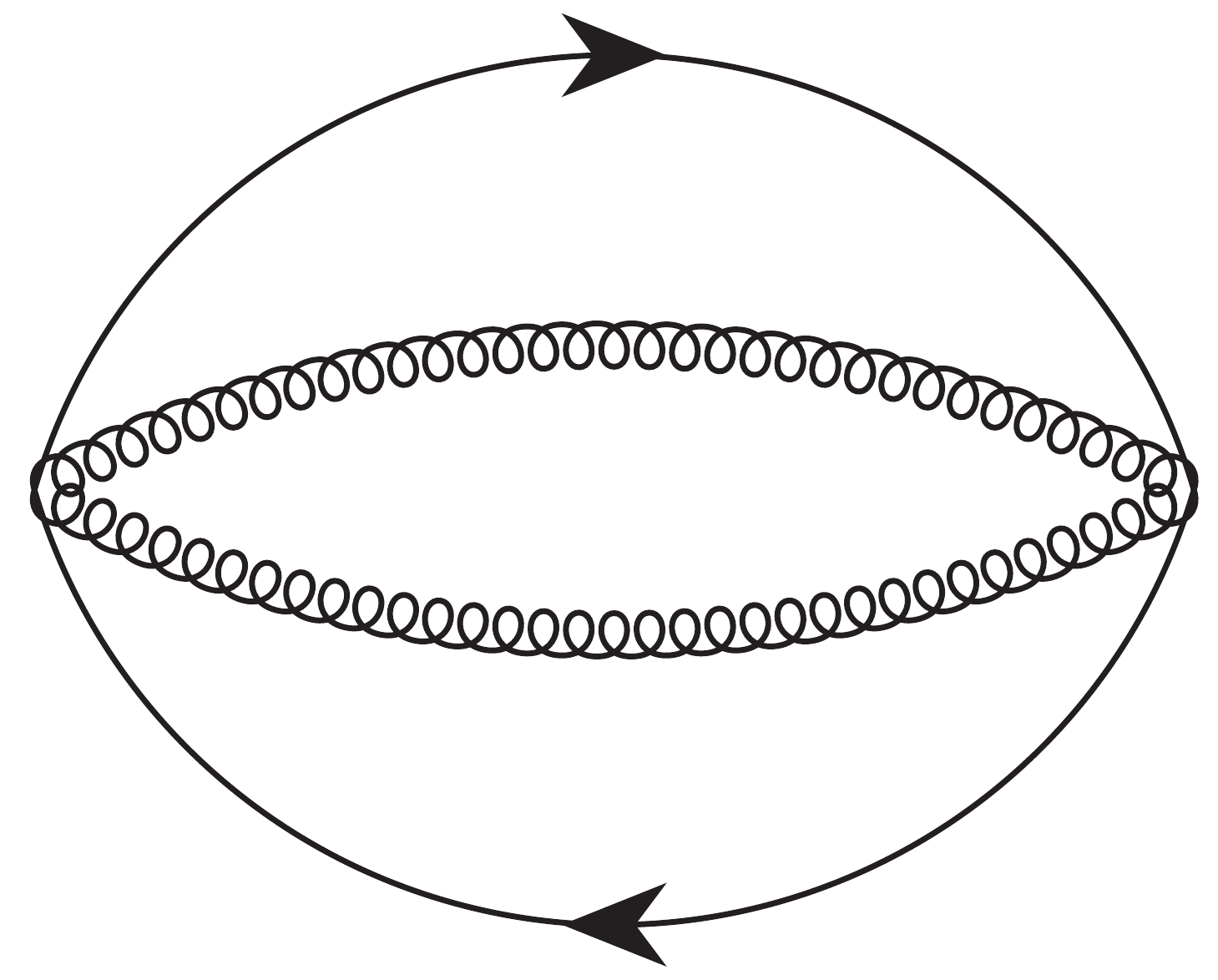}}}
\\
\subfigure[(b--1)]{
\scalebox{0.12}{\includegraphics{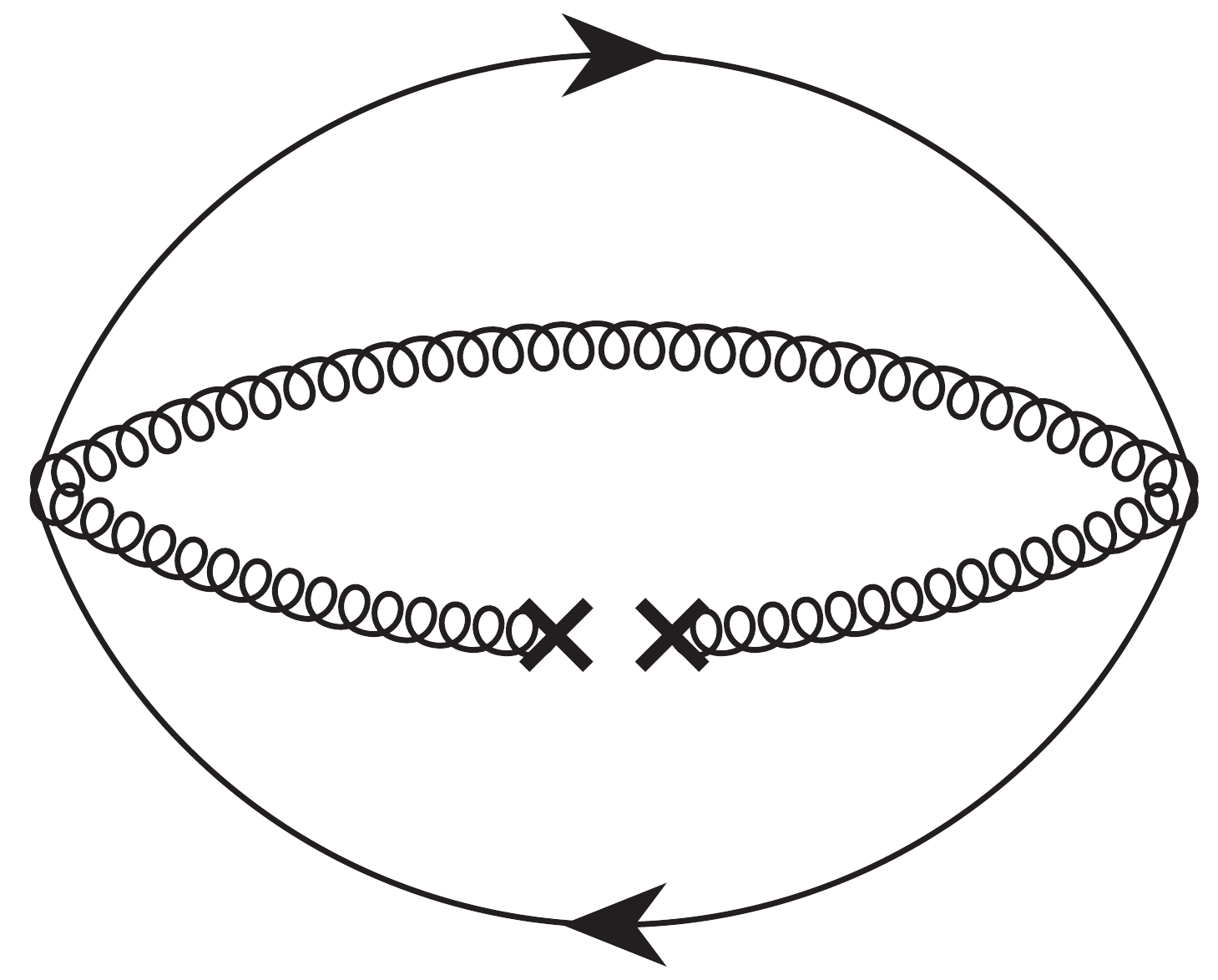}}}~
\subfigure[(b--2)]{
\scalebox{0.12}{\includegraphics{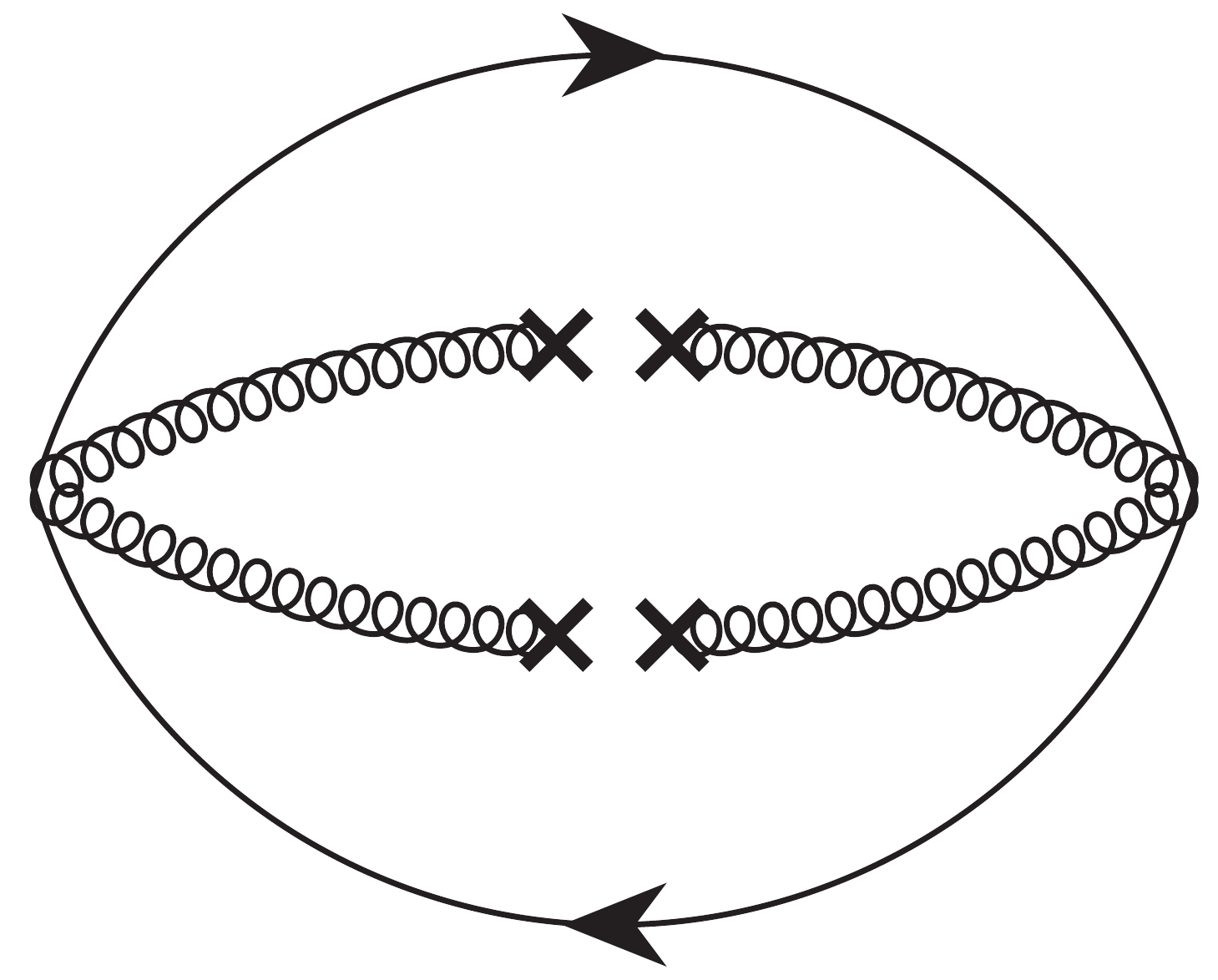}}}~
\subfigure[(b--3)]{
\scalebox{0.12}{\includegraphics{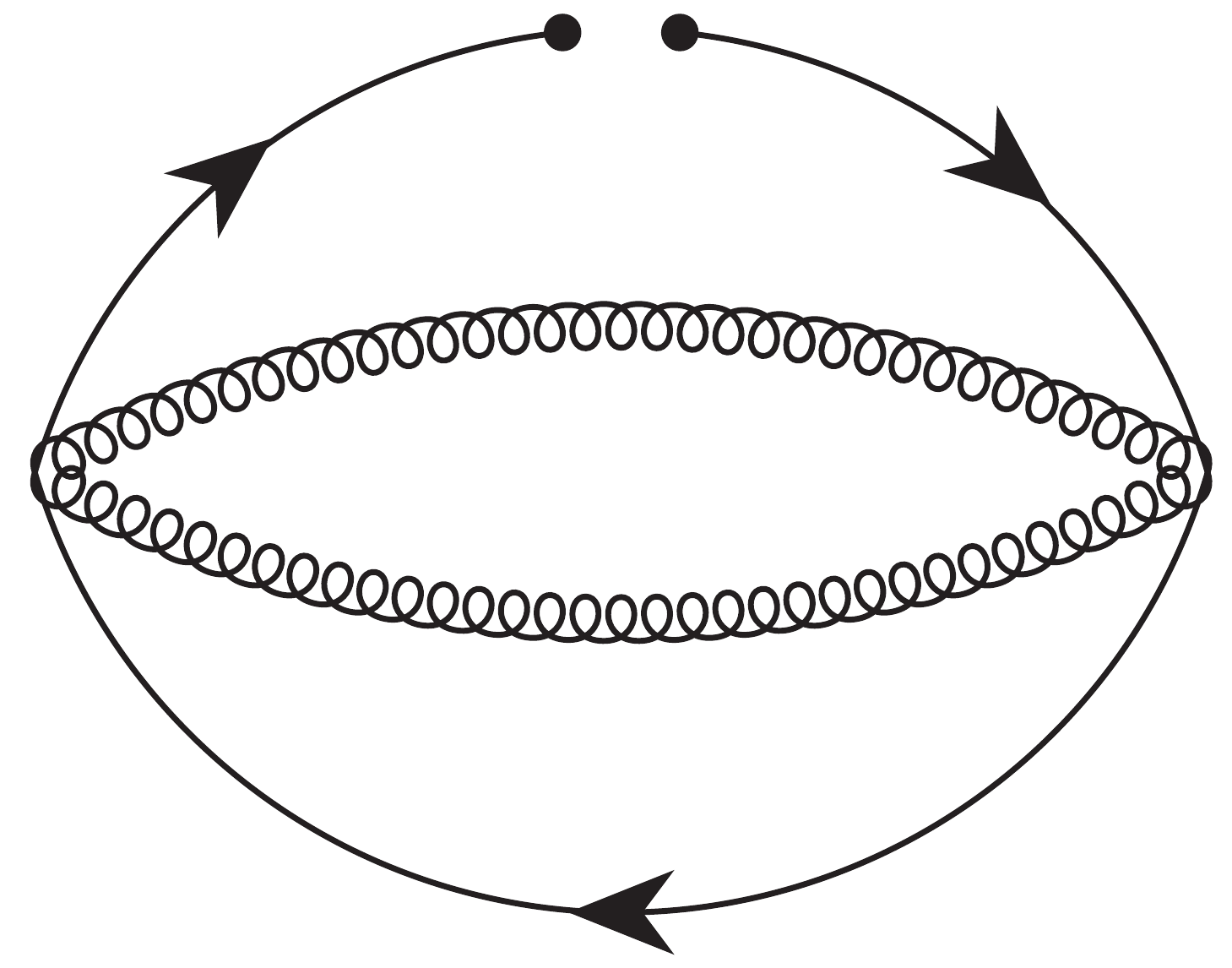}}}~
\subfigure[(b--4)]{
\scalebox{0.12}{\includegraphics{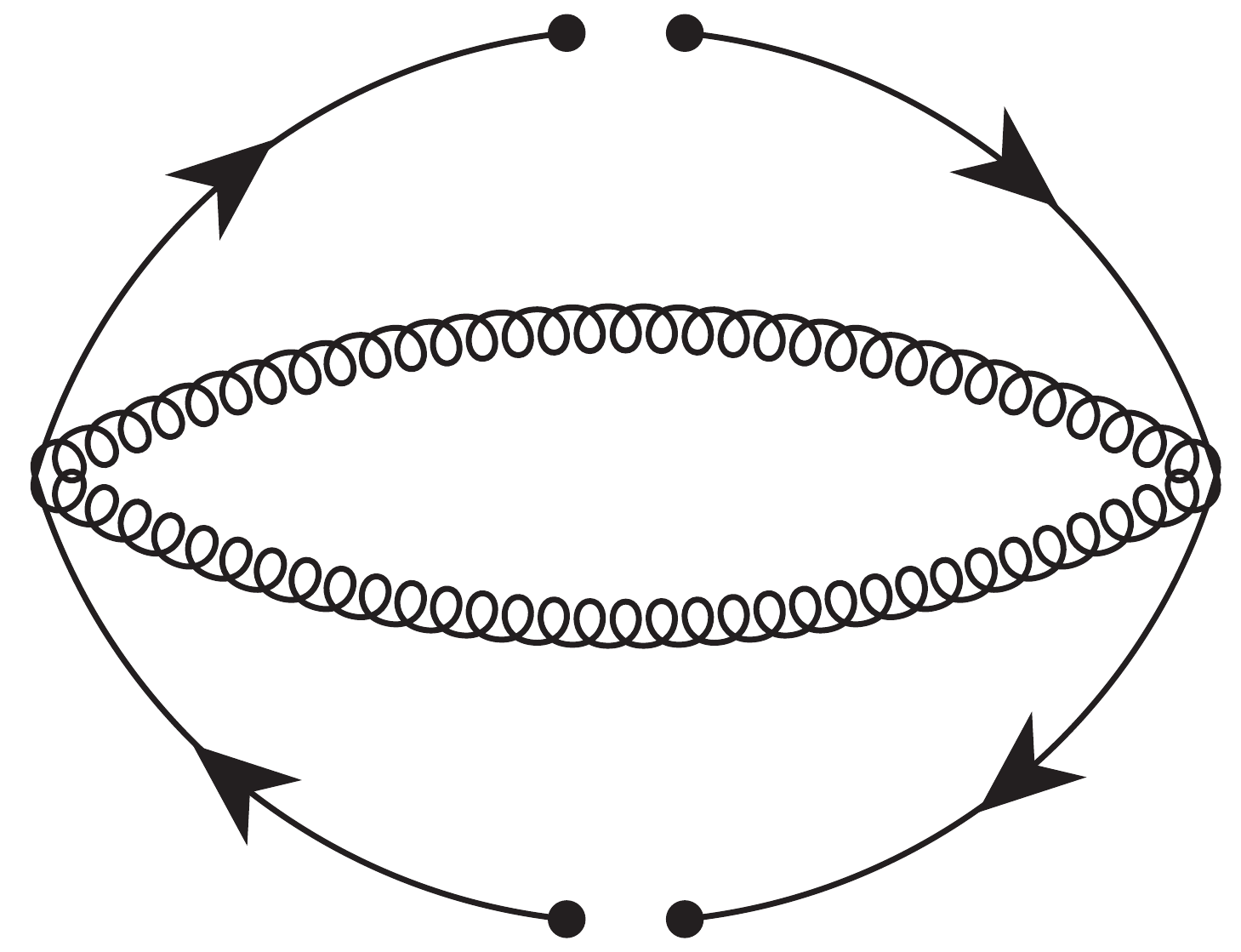}}}
\\
\subfigure[(c--1)]{
\scalebox{0.12}{\includegraphics{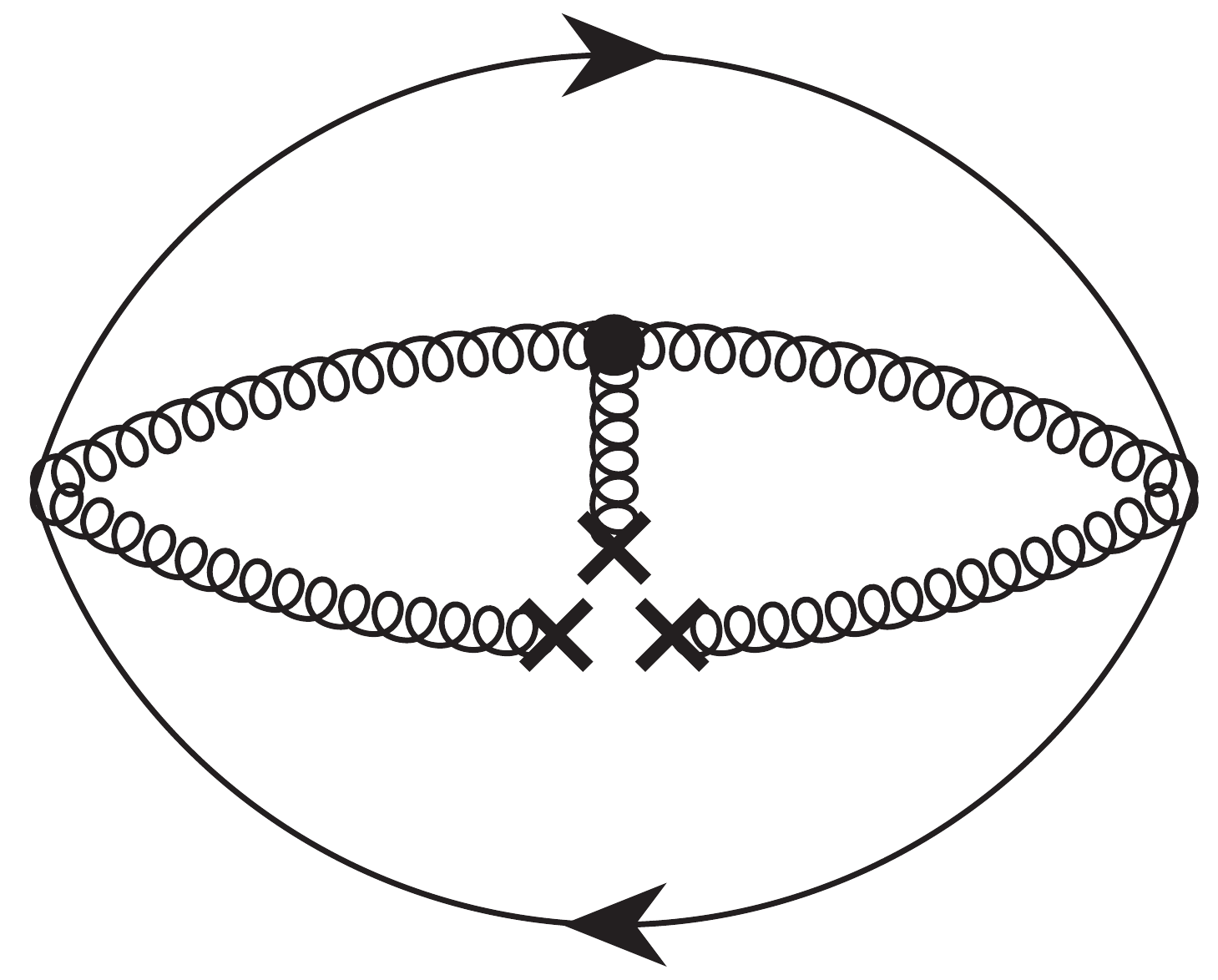}}}~
\subfigure[(c--2)]{
\scalebox{0.12}{\includegraphics{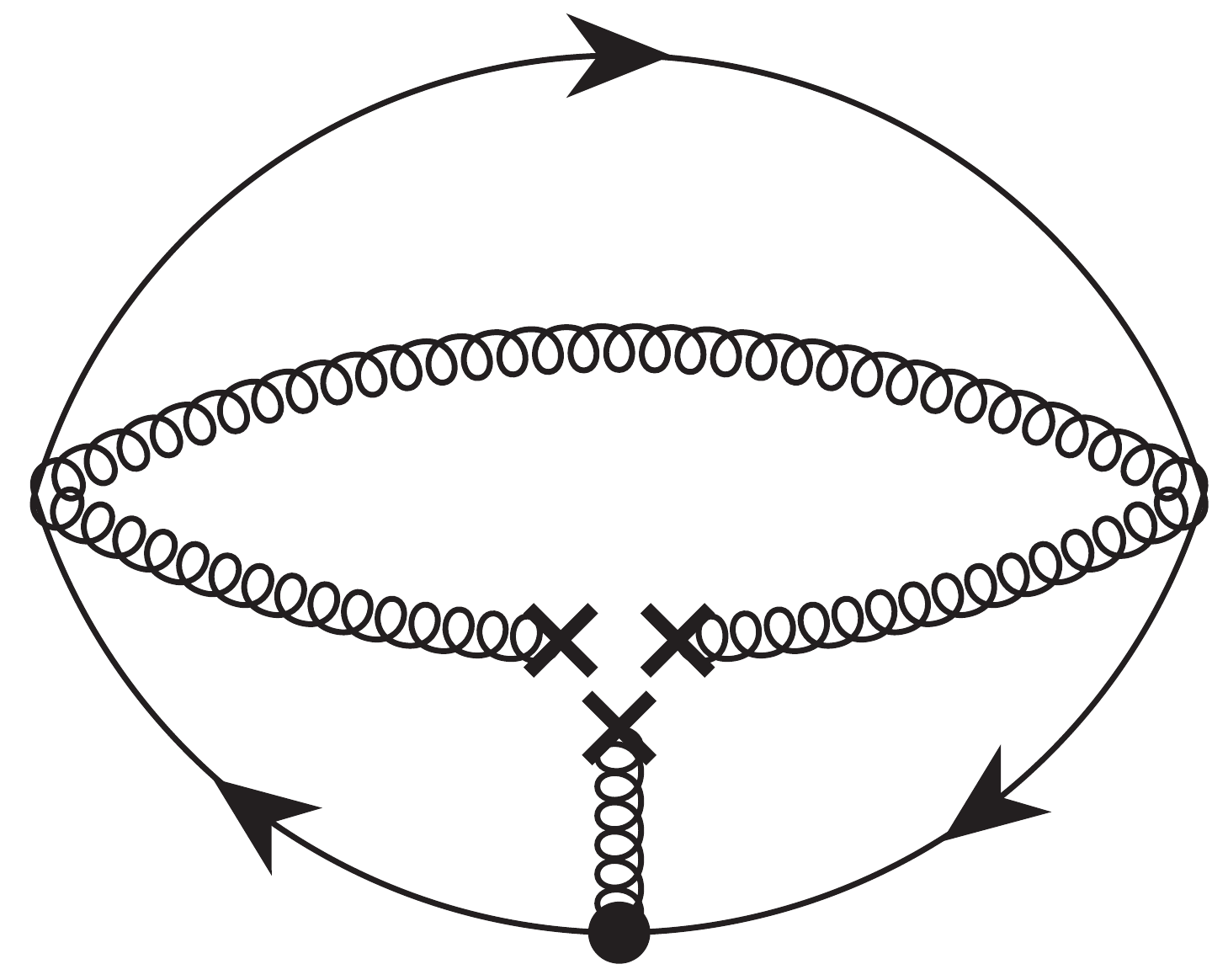}}}~
\subfigure[(c--3)]{
\scalebox{0.12}{\includegraphics{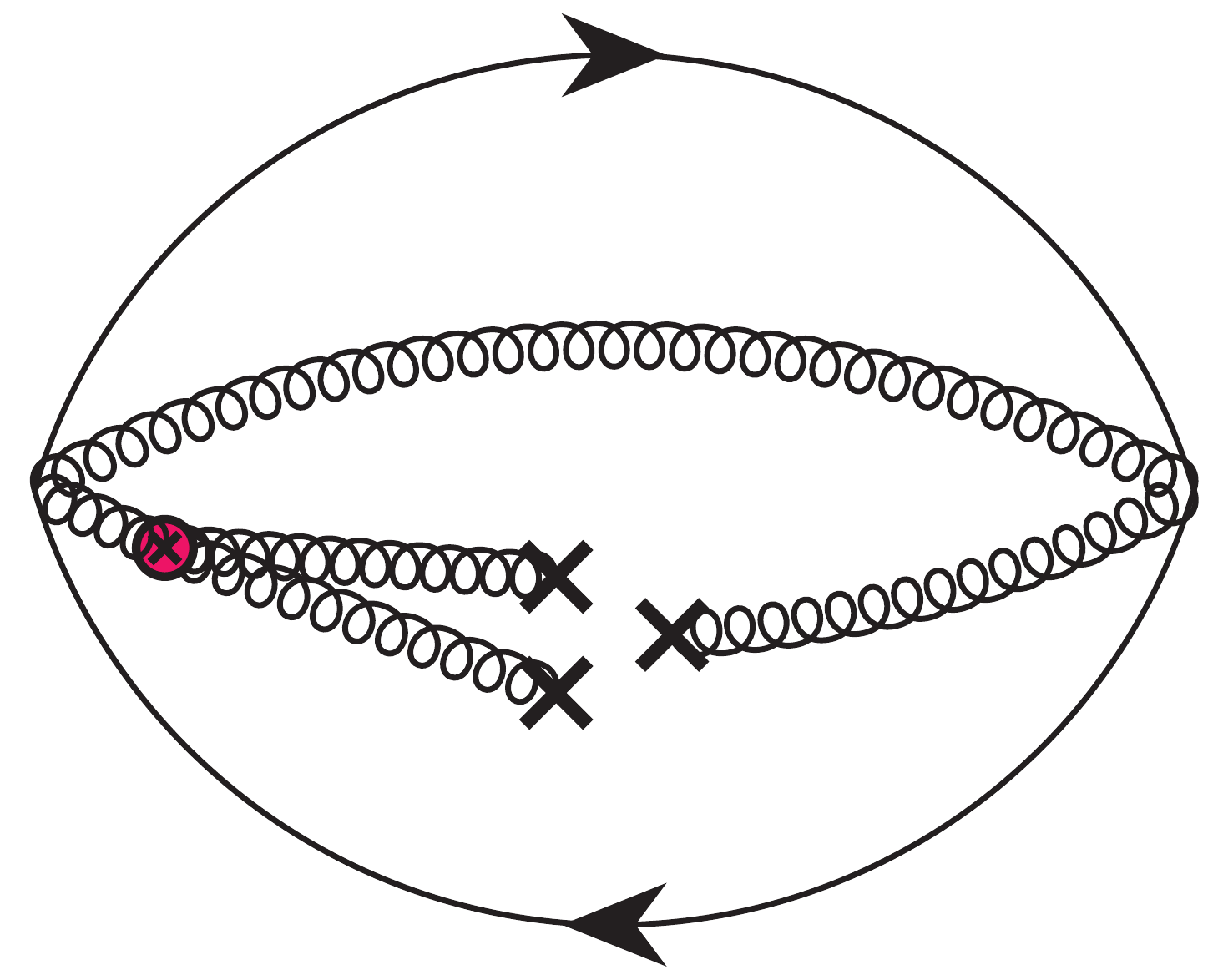}}}~
\subfigure[(c--4)]{
\scalebox{0.12}{\includegraphics{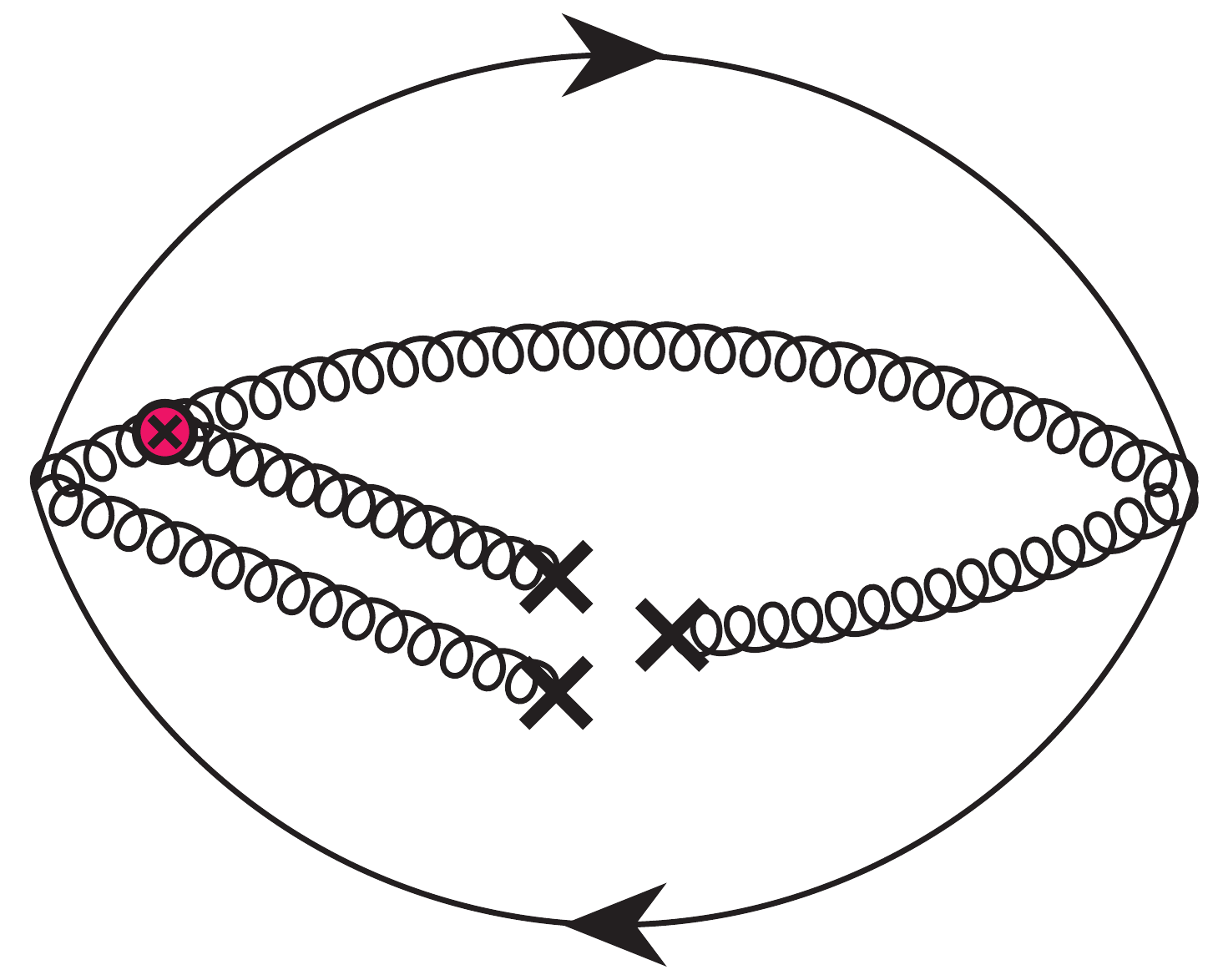}}}
\\
\subfigure[(d--1)]{
\scalebox{0.12}{\includegraphics{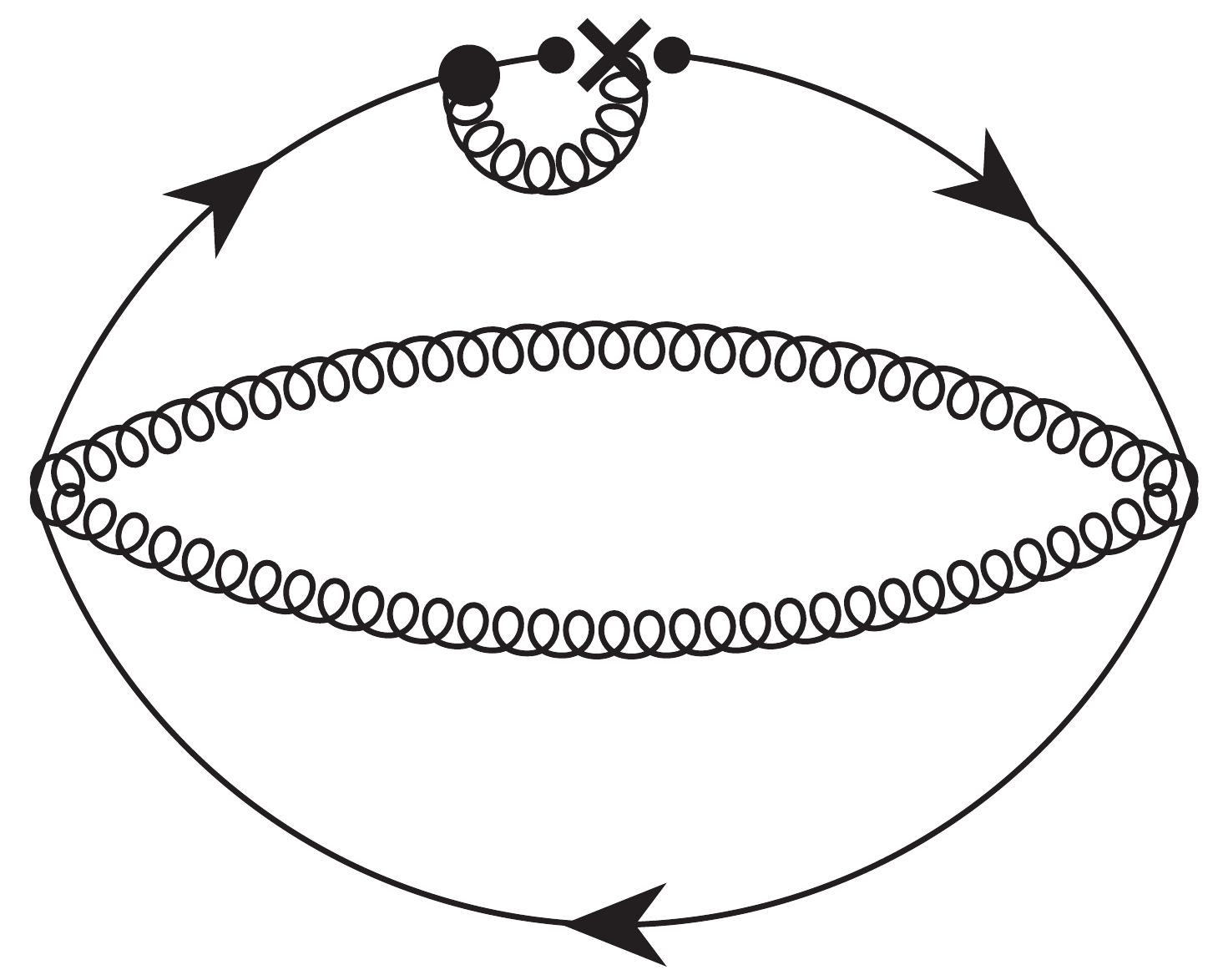}}}~
\subfigure[(d--2)]{
\scalebox{0.12}{\includegraphics{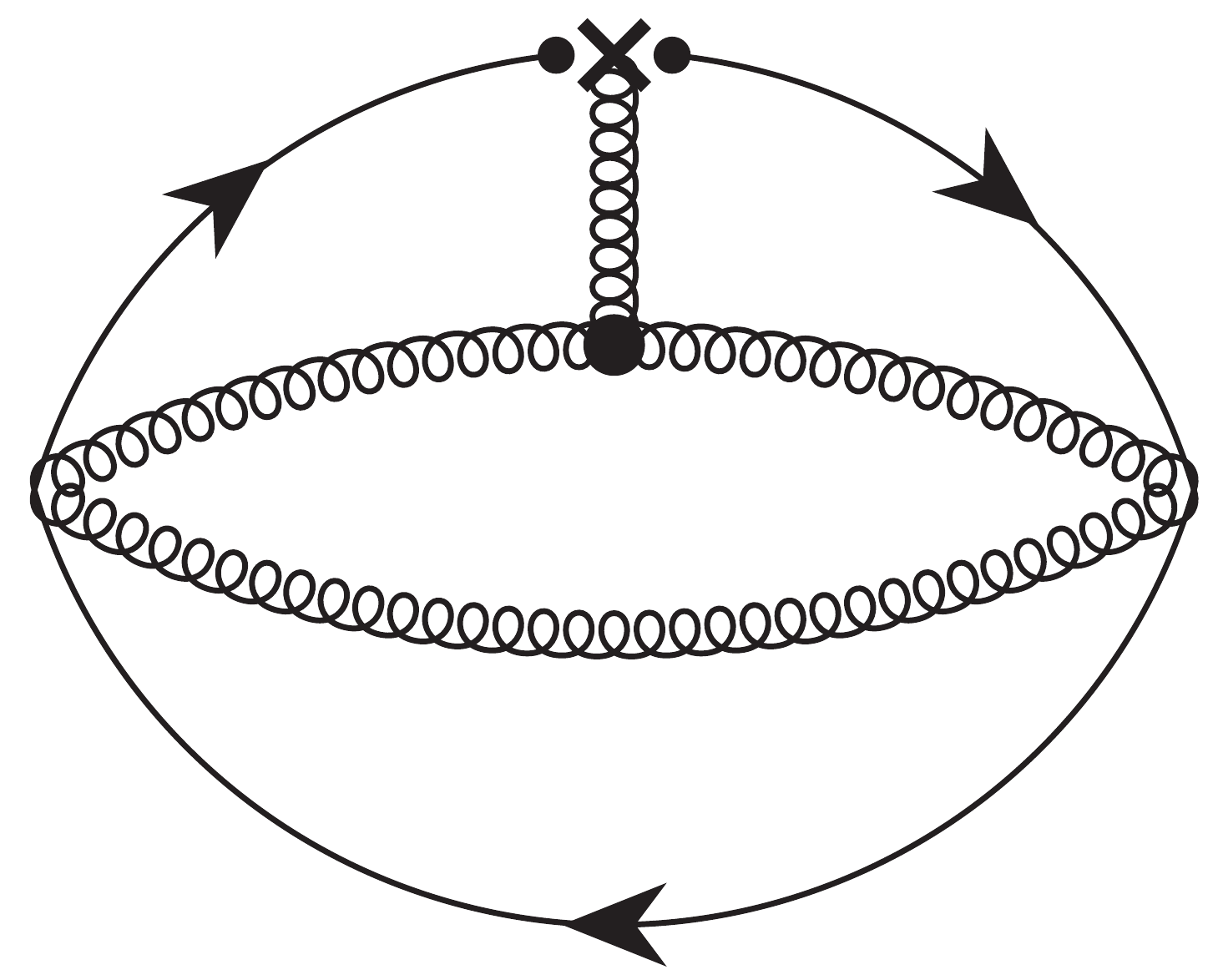}}}~
\subfigure[(d--3)]{
\scalebox{0.12}{\includegraphics{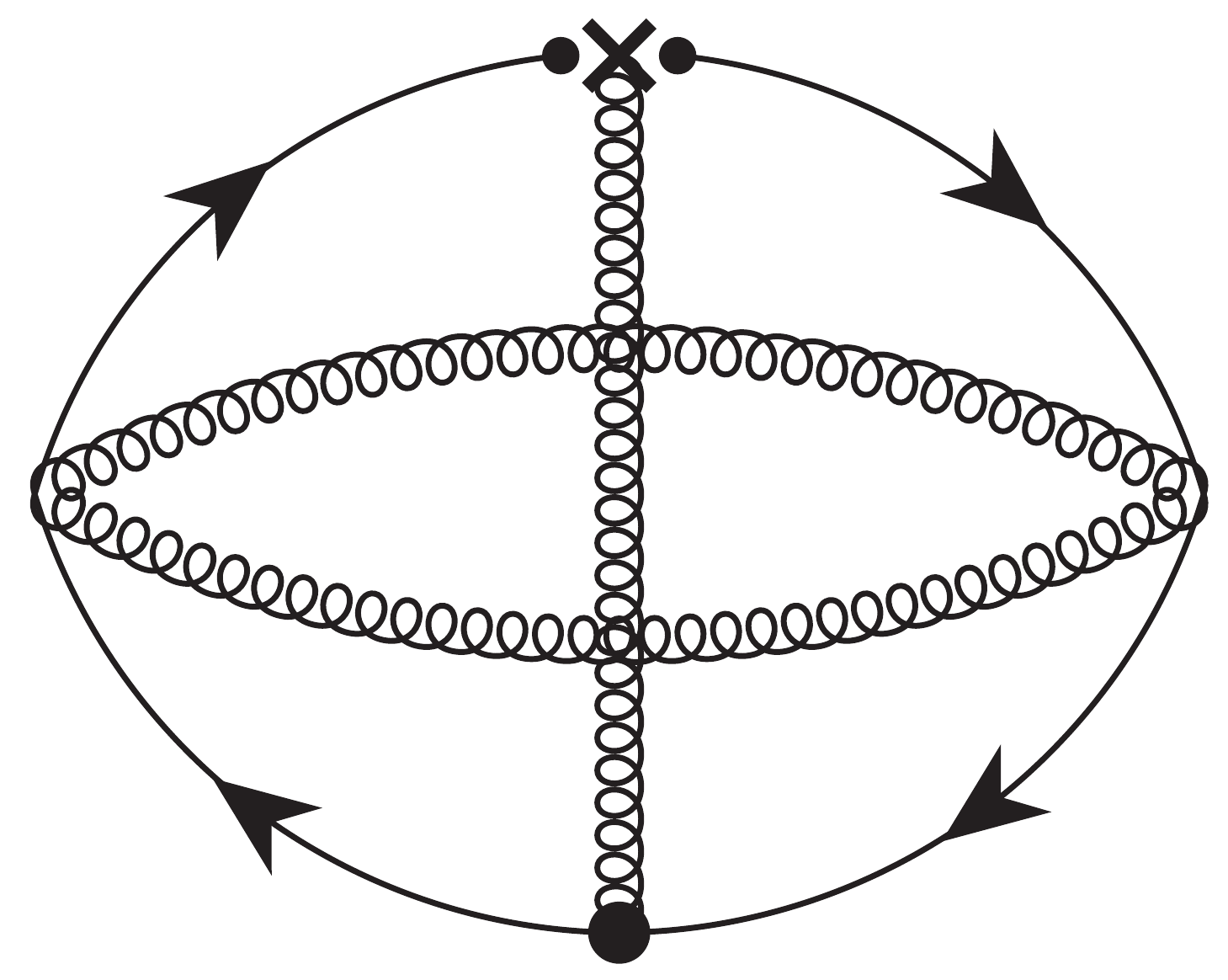}}}
\\
\subfigure[(d--4)]{
\scalebox{0.12}{\includegraphics{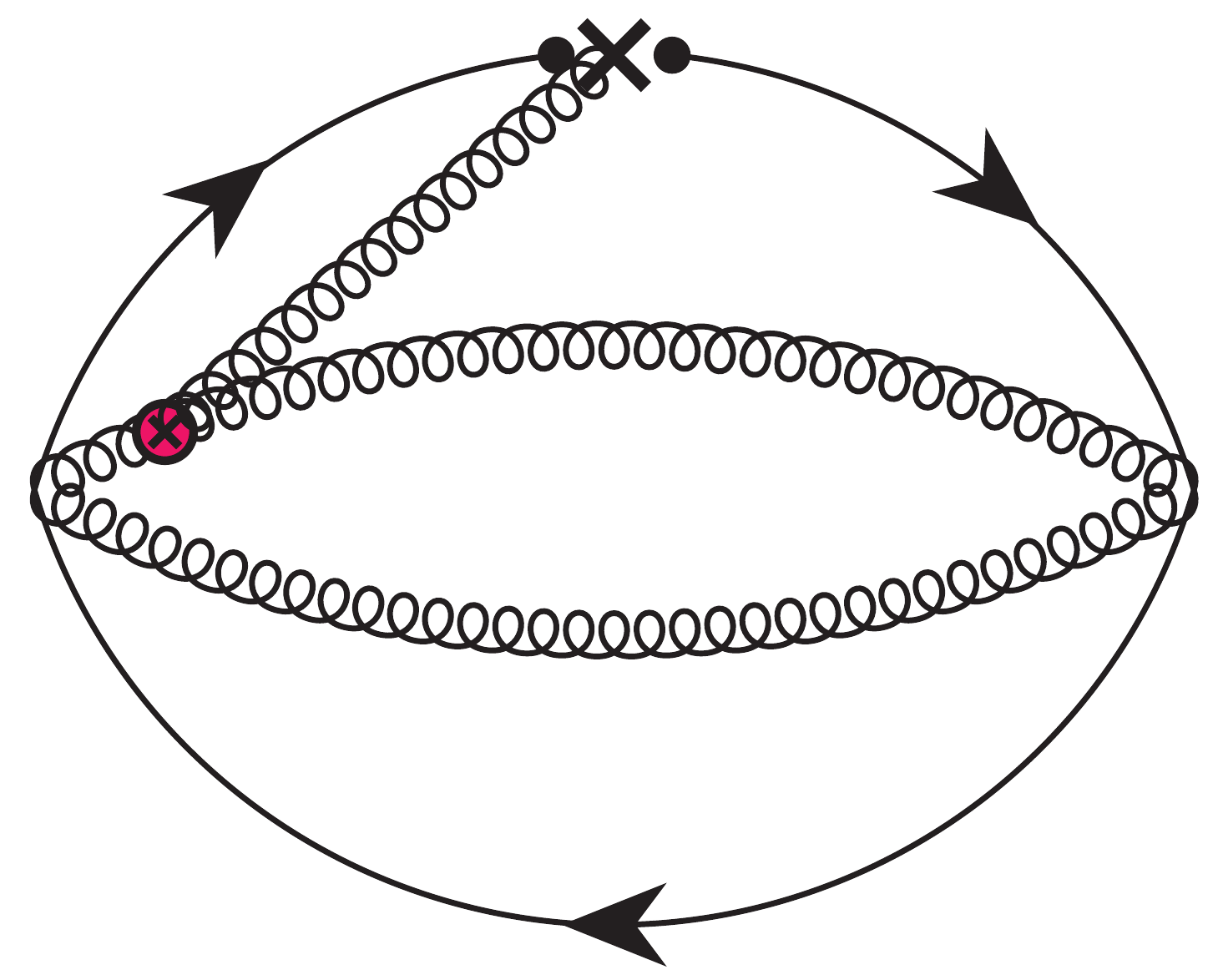}}}~
\subfigure[(d--5)]{
\scalebox{0.12}{\includegraphics{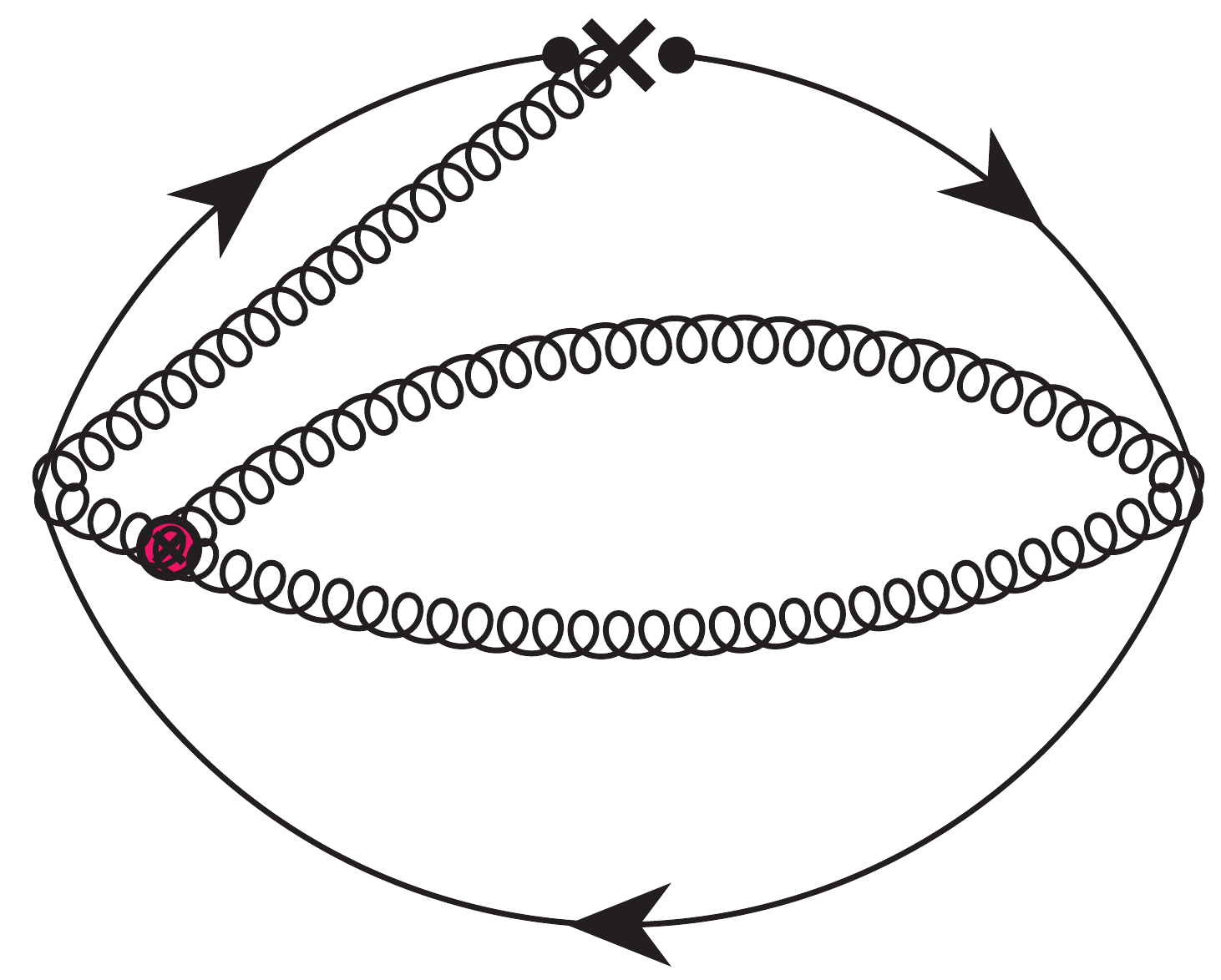}}}~
\subfigure[(d--6)]{
\scalebox{0.12}{\includegraphics{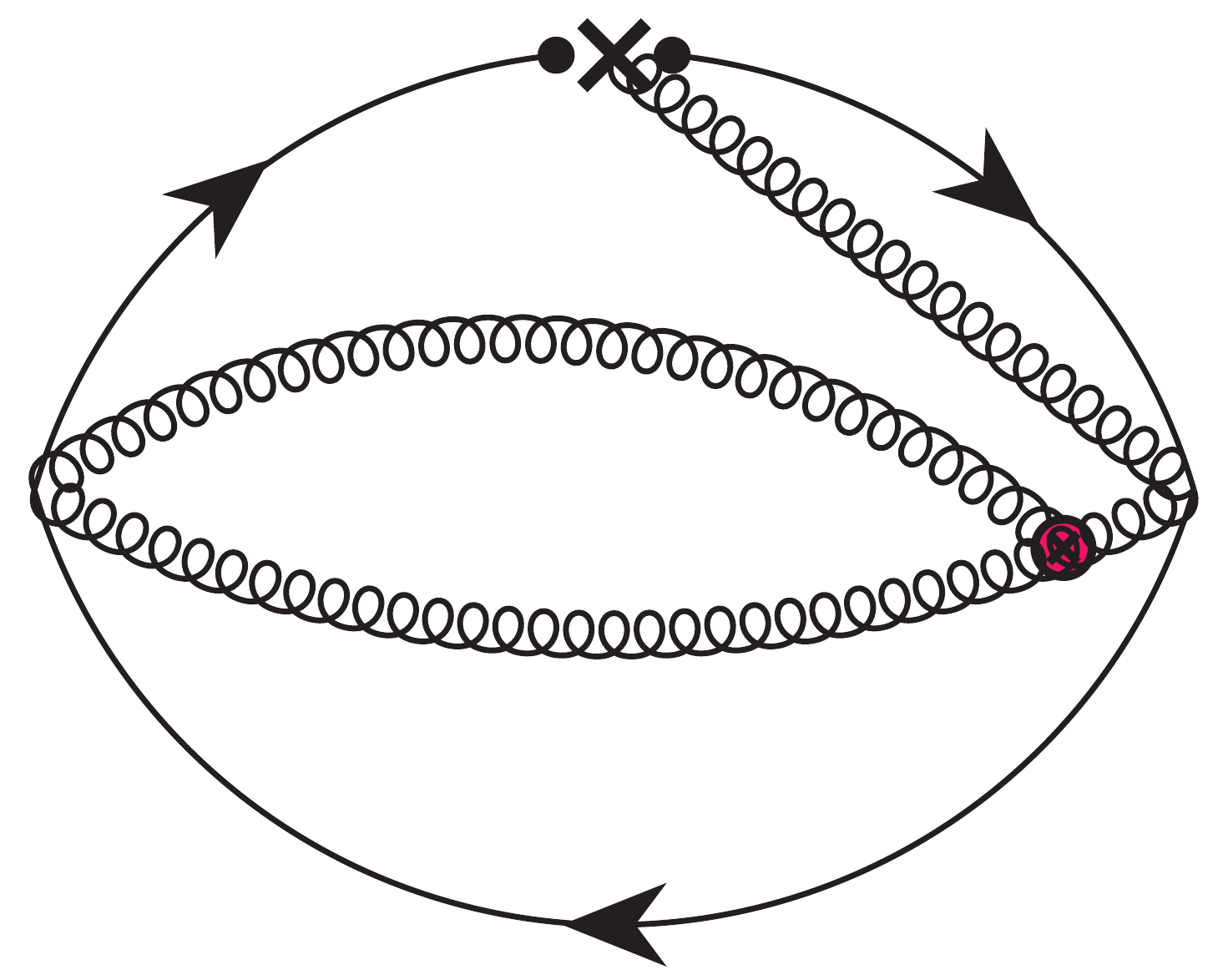}}}
\\
\subfigure[(e--1)]{
\scalebox{0.12}{\includegraphics{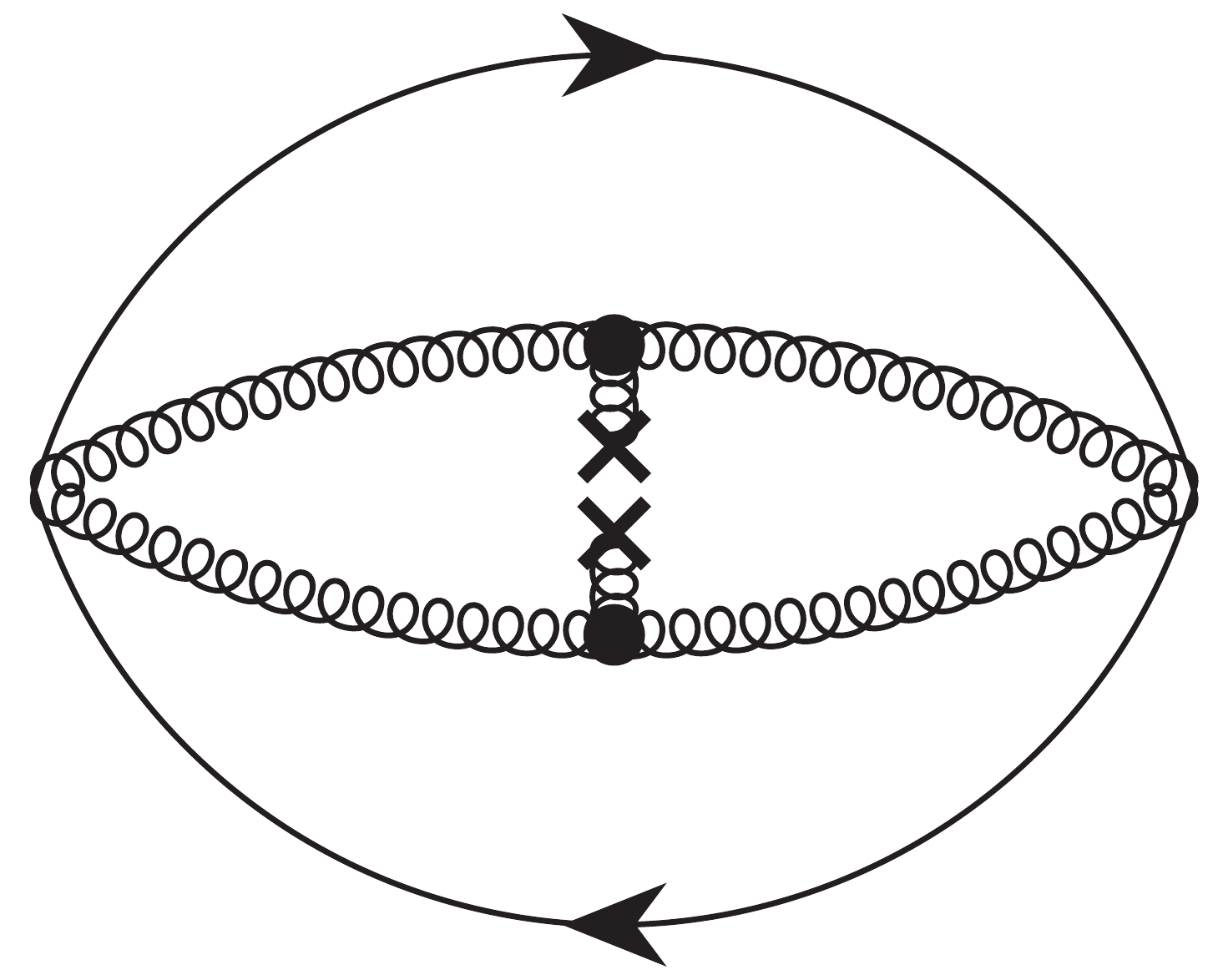}}}~
\subfigure[(e--2)]{
\scalebox{0.12}{\includegraphics{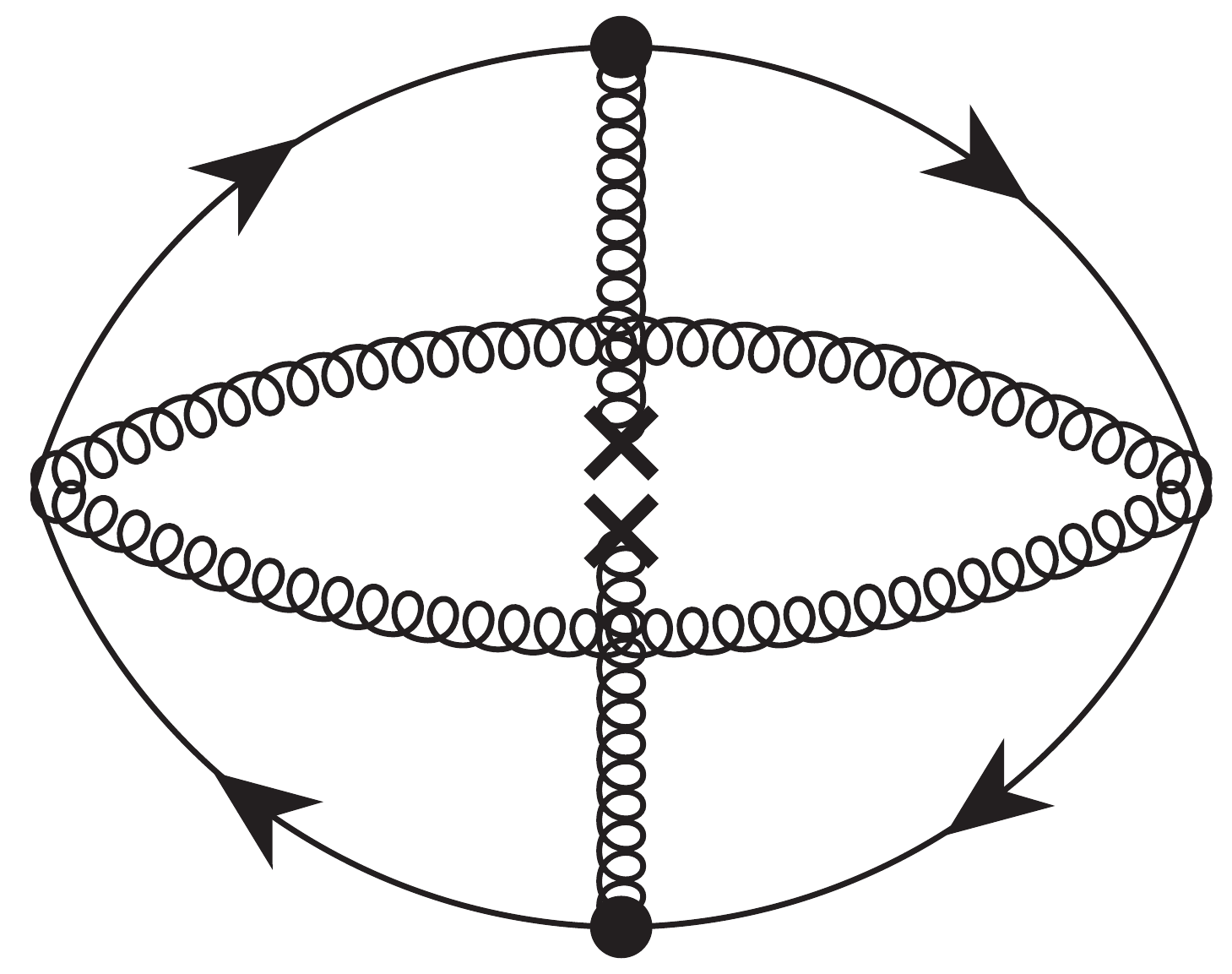}}}~
\subfigure[(e--3)]{
\scalebox{0.12}{\includegraphics{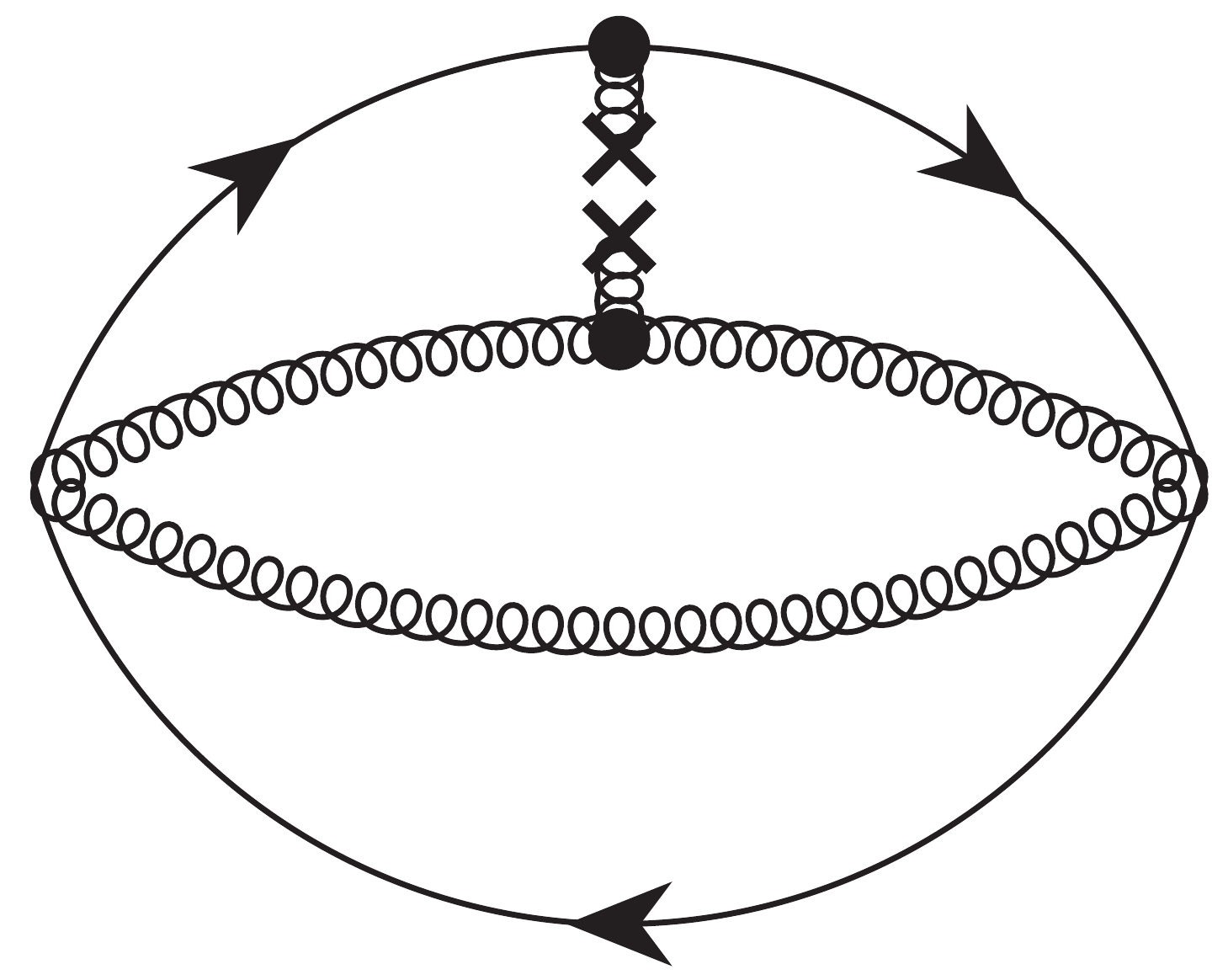}}}
\end{center}
\caption{Feynman diagrams for the double-gluon hybrid state: (a) and (b--i) are proportional to $\alpha_s^2 \times g_s^0$; (c--i) and (d--i) are proportional to $\alpha_s^2 \times g_s^1$; (e--i) are proportional to $\alpha_s^2 \times g_s^2$.}
\label{fig:feynman}
\end{figure}

In this paper we have calculated the Feynman diagrams depicted in Fig.~\ref{fig:feynman}. Since the gluon field strength tensor $G^n_{\mu\nu}$ is defined as
\begin{equation}
G^n_{\mu\nu} = \partial_\mu A_\nu^n  -  \partial_\nu A_\mu^n  +  g_s f^{npq} A_{p,\mu} A_{q,\nu} \, ,
\end{equation}
it is naturally separated into two parts: the former two terms are represented by the single-gluon-line, and the third term is represented by the double-gluon-line with a red vertex, {\it e.g.}, see the diagram depicted in Fig.~\ref{fig:feynman}(c--3).

We have calculated $\rho_{\rm OPE}(s)$ up to the dimension eight condensates, including the perturbative term, the quark condensates, the quark-gluon mixed condensates, the two-/three-gluon condensates, and their combinations. We have considered all the diagrams proportional to $\alpha_s^2 \times g_s^0$ and $\alpha_s^2 \times g_s^1$, while we have only considered three diagrams proportional to $\alpha_s^2 \times g_s^2$, as depicted in Fig.~\ref{fig:feynman}(e--i). We have included the strange quark mass, while we have neglected the up and down quark masses.

All the obtained spectral densities are given in Appendix~\ref{app:sumrule}. Especially, the one extracted from the current $J_{0^{++}}$ with the quark-gluon content $\bar q q gg$ ($q=u/d$) is
\begin{eqnarray}
\rho^{\bar q q gg}_{0^{++}}(s) &=& \frac{\alpha_s^2 s^5}{4320 \pi^4} - \frac{5\alpha_s^2 \langle g_s^2 GG \rangle s^3}{6912 \pi^4}
\label{rho:0pp}
\\ \nonumber &+& \left( \frac{80 \alpha_s^2 \langle \bar q q \rangle^2}{27} - \frac{5 \alpha_s \langle g_s^3 G^3 \rangle}{48\pi^3} \right) s^2
\\ \nonumber &+& \left( - \frac{80\alpha_s^2 \langle \bar q q \rangle \langle g_s \bar q \sigma G q \rangle}{9} - \frac{5 \langle g_s^2 GG \rangle^2}{288\pi^2} \right) s \, ,
\end{eqnarray}
and the one with the quark-gluon content $\bar s s gg$ is
\begin{eqnarray}
\rho^{\bar s s gg}_{0^{++}}(s) &=& \frac{\alpha_s^2 s^5}{4320 \pi^4}- \frac{\alpha_s^2 m_s^2 s^4}{144 \pi^4}
\label{rho:0ppss}
\\ \nonumber &+&
\left( - \frac{5\alpha_s^2 \langle g_s^2 GG \rangle }{6912 \pi^4}
- \frac{5\alpha_s^2 m_s \langle \bar s s \rangle }{27 \pi^2}
+ \frac{5\alpha_s^2 m_s^4 }{36 \pi^4} \right) s^3
\\ \nonumber &+&
\left(
\frac{80 \alpha_s^2 \langle \bar s s \rangle^2}{27}
- \frac{5 \alpha_s \langle g_s^3 G^3 \rangle}{48\pi^3}
\right.
\\ \nonumber &&
\left.
+\frac{10 \alpha_s^2 m_s  \langle g_s \bar s \sigma G s \rangle}{9\pi^2}
-\frac{20 \alpha_s^2 m_s^3  \langle \bar s s \rangle}{9\pi^2}
\right) s^2
\\ \nonumber &+&
\left( - \frac{80\alpha_s^2 \langle \bar s s \rangle \langle g_s \bar s \sigma G s \rangle}{9} \right.
- \frac{5 \langle g_s^2 GG \rangle^2}{288\pi^2}
\\ \nonumber &&
\left.
+\frac{40 \alpha_s^2 m_s^2 \langle \bar s s \rangle^2}{9}
+\frac{5 \alpha_s m_s^2 \langle g_s^3 G^3 \rangle}{8\pi^3}
\right) s \, .
\end{eqnarray}
Note that the double-gluon hybrid states in the same isospin multiplet have the same extracted hadron mass within our QCD sum rule framework, since we do not differentiate the up and down quarks in the OPE series.

\section{Numerical Analyses}
\label{sec:numerical}
%

%
\begin{figure*}[hbt]
\begin{center}
\includegraphics[width=0.45\textwidth]{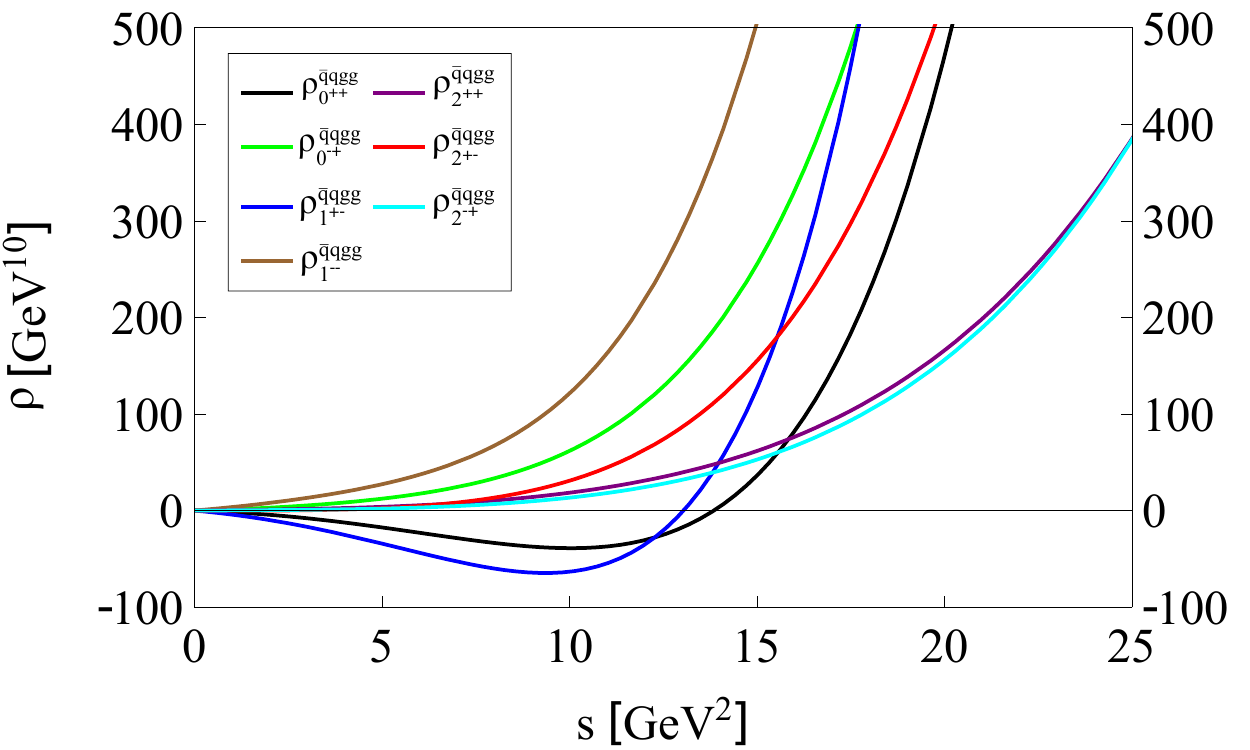}
~~~~~
\includegraphics[width=0.45\textwidth]{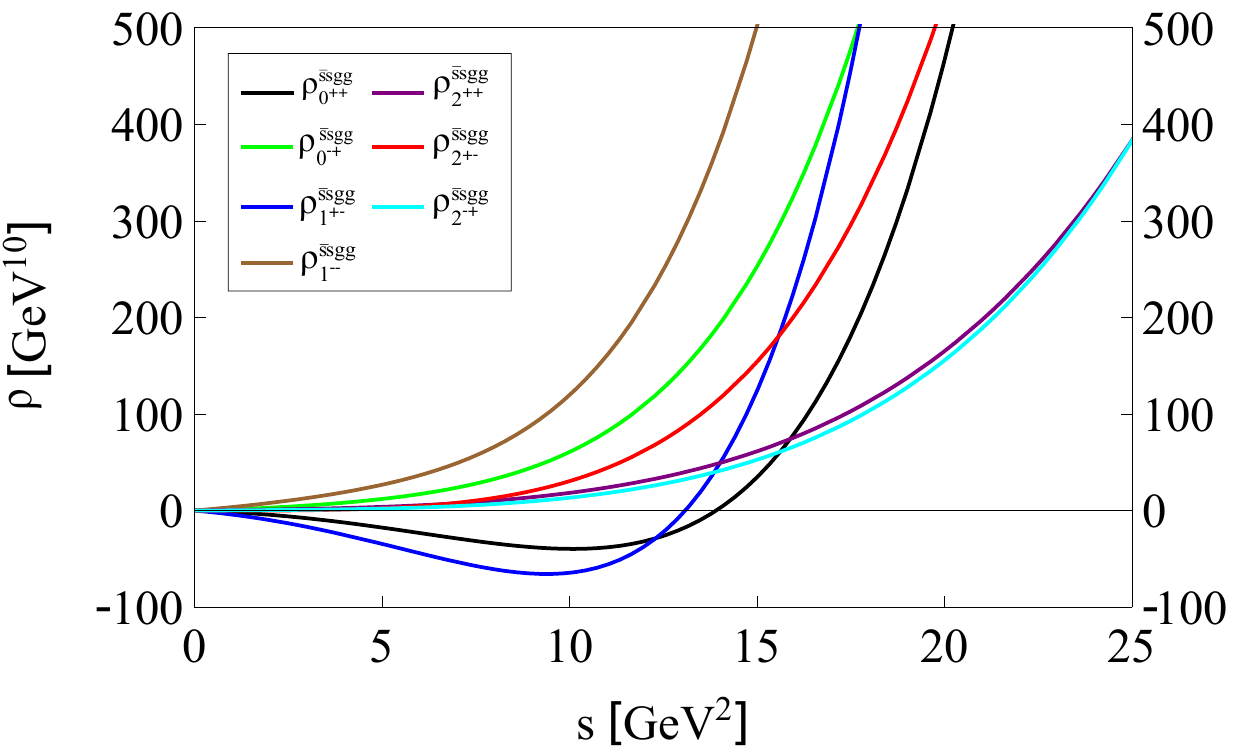}
\caption{The spectral densities $\rho^{\bar q q gg}_{J^{PC}}(s)$ and $\rho^{\bar s s gg}_{J^{PC}}(s)$ as functions of $s$.}
\label{fig:pi12}
\end{center}
\end{figure*}

In this section we use the spectral densities listed in Eq.~(\ref{rho:0pp}) and Eq.~(\ref{rho:0ppss}) as well as those given in Appendix~\ref{app:sumrule} to perform numerical analyses. To begin with, we show them in Fig.~\ref{fig:pi12} as functions of $s$. The two spectral densities $\rho^{\bar q q gg}_{0^{++}}(s)$ and $\rho^{\bar s s gg}_{0^{++}}(s)$ (black curves) are both negative when $0 \leq s \leq 14$~GeV$^2$. This suggests that they are both non-physical in this energy region, and the masses extracted from them should be significantly larger than $\sqrt{14}$~GeV~$\approx 4$~GeV. However, the two spectral densities $\rho^{\bar q q gg}_{2^{+-}}(s)$ and $\rho^{\bar s s gg}_{2^{+-}}(s)$ (red curves) are both positive definite, and the masses extracted from them can be much smaller.

We use the spectral density $\rho^{\bar q q gg}_{0^{++}}(s)$ listed in Eq.~(\ref{rho:0pp}) as an example. It is extracted from the current $J_{0^{++}}$ with the quark-gluon content $\bar q q gg$ ($q=u/d$), so we denote its corresponding state as
\begin{equation}
X \equiv | X; 0^{++} \rangle \equiv |\bar q q gg; 0^{++}\rangle \, .
\end{equation}
The following values will be used for various QCD parameters at the renormalization scale 2 GeV and the QCD scale $\Lambda_{\rm QCD} = 300$~MeV~\cite{pdg,Ovchinnikov:1988gk,Yang:1993bp,Ellis:1996xc,Ioffe:2002be,Jamin:2002ev,Gimenez:2005nt,Narison:2011xe,Narison:2018dcr}:
\begin{eqnarray}
\nonumber \alpha_s(Q^2) &=& {4\pi \over 11 \ln(Q^2/\Lambda_{\rm QCD}^2)} \, ,
\\ \nonumber m_s &=& 93 ^{+11}_{-5} \mbox{ MeV} \, ,
\\ \nonumber \langle\bar qq \rangle &=& -(0.240 \pm 0.010)^3 \mbox{ GeV}^3 \, ,
\\ \langle\bar ss \rangle &=& (0.8\pm 0.1)\times \langle\bar qq \rangle \, ,
\label{eq:condensate}
\\  \nonumber \langle g_s\bar q\sigma G q\rangle &=& (0.8 \pm 0.2)\times\langle\bar qq\rangle \mbox{ GeV}^2 \, ,
\\ \nonumber \langle g_s\bar s\sigma G s\rangle &=&  (0.8 \pm 0.2)\times\langle\bar ss\rangle \, ,
\\ \nonumber \langle \alpha_s GG\rangle &=& (6.35 \pm 0.35) \times 10^{-2} \mbox{ GeV}^4 \, ,
\\ \nonumber \langle g_s^3G^3\rangle &=& (8.2 \pm 1.0) \times \langle \alpha_s GG\rangle  \mbox{ GeV}^2 \, .
\end{eqnarray}

Eq.~(\ref{eq:LSR}) indicates that the mass $M_X$ depends on two free parameters: the threshold value $s_0$ and the Borel mass $M_B$. We use three criteria to determine their proper working regions: a) the convergence of OPE is sufficiently good, b) the pole contribution is sufficiently large, and c) the mass dependence on these two parameters is sufficiently weak.

\begin{figure}[hbtp]
\begin{center}
\includegraphics[width=0.45\textwidth]{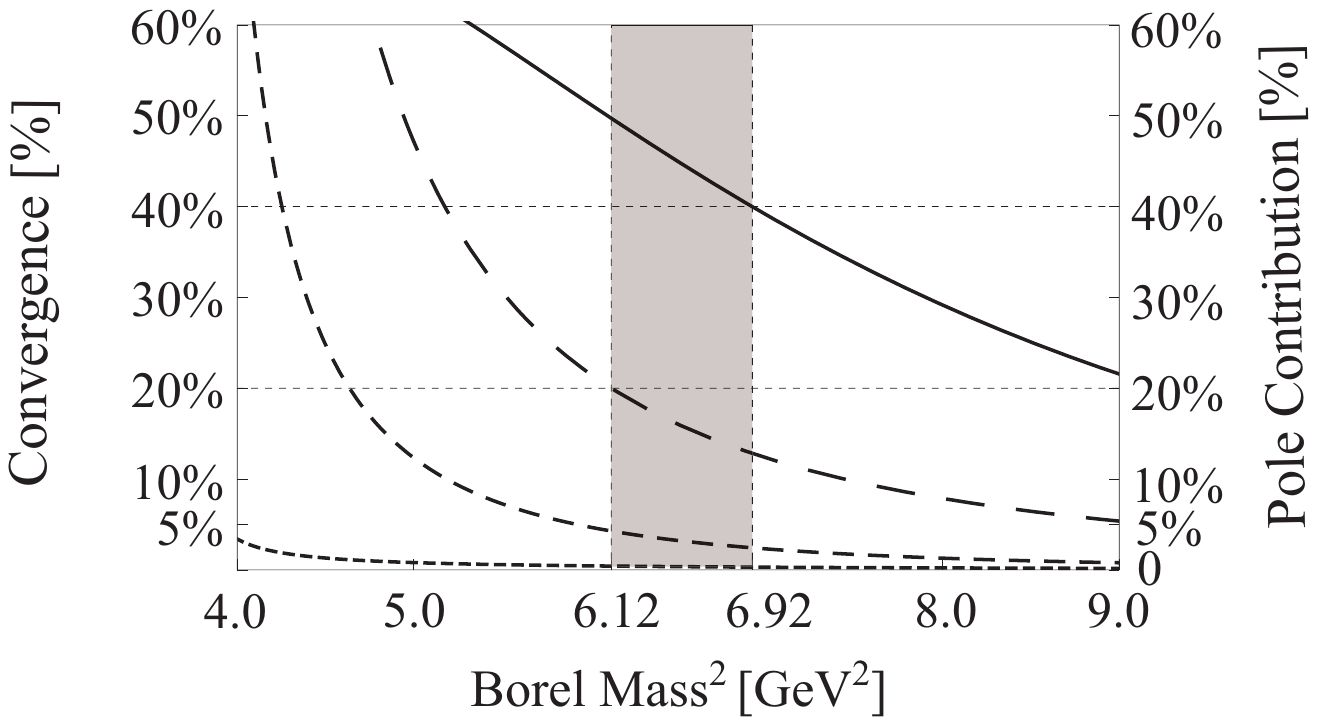}
\caption{CVG (short-dashed curve, defined in Eq.~(\ref{eq:convergence1})), CVG$^\prime$ (middle-dashed curve, defined in Eq.~(\ref{eq:convergence2})), CVG$^{\prime\prime}$ (long-dashed curve, defined in Eq.~(\ref{eq:convergence3})), and PC (solid curve, defined in Eq.~(\ref{eq:pole})) as functions of the Borel mass $M_B$, when setting $s_0 = 38.0$~GeV$^2$. These curves are obtained using the spectral density $\rho^{\bar q q gg}_{0^{++}}(s)$ extracted from the current $J_{0^{++}}$ with the quark-gluon content $\bar q q gg$ ($q=u/d$).}
\label{fig:cvgpole}
\end{center}
\end{figure}

In order to ensure the good convergence of OPE, we require that the $\alpha_s^2 \times g_s^2$ terms are less than 5\%, the $D=8$ terms are less than 10\%, and the $D=6$ terms are less than 20\%:
\begin{eqnarray}
\mbox{CVG} &\equiv& \left|\frac{ \Pi^{g_s^{n=6}}(\infty, M_B^2) }{ \Pi(\infty, M_B^2) }\right| \leq 5\% \, ,\label{eq:convergence1}
\\
\mbox{CVG}^\prime &\equiv& \left|\frac{ \Pi^{{\rm D=8}}(\infty, M_B^2) }{ \Pi(\infty, M_B^2) }\right| \leq 10\% \, ,\label{eq:convergence2}
\\
\mbox{CVG}^{\prime\prime} &\equiv& \left|\frac{ \Pi_{11}^{D=6}(\infty, M_B^2) }{ \Pi_{11}(\infty, M_B^2) }\right| \leq 20\% \, .\label{eq:convergence3}
\end{eqnarray}
As depicted in Fig.~\ref{fig:cvgpole} using three dashed curves, we determine the minimum Borel mass to be $M_B^2 \geq 6.12$~GeV$^2$.

Then we require the pole contribution (PC) to be sufficiently large, that is larger than $40\%$:
\begin{equation}
\mbox{PC} \equiv \left|\frac{ \Pi(s_0, M_B^2) }{ \Pi(\infty, M_B^2) }\right| \geq 40\% \, .
\label{eq:pole}
\end{equation}
As depicted in Fig.~\ref{fig:cvgpole} using the solid curve, we determine the maximum Borel mass to be $M_B^2 \leq 6.92$~GeV$^2$, when setting $s_0 = 38.0$~GeV$^2$.

\begin{figure*}[hbtp]
\begin{center}
\includegraphics[width=0.4\textwidth]{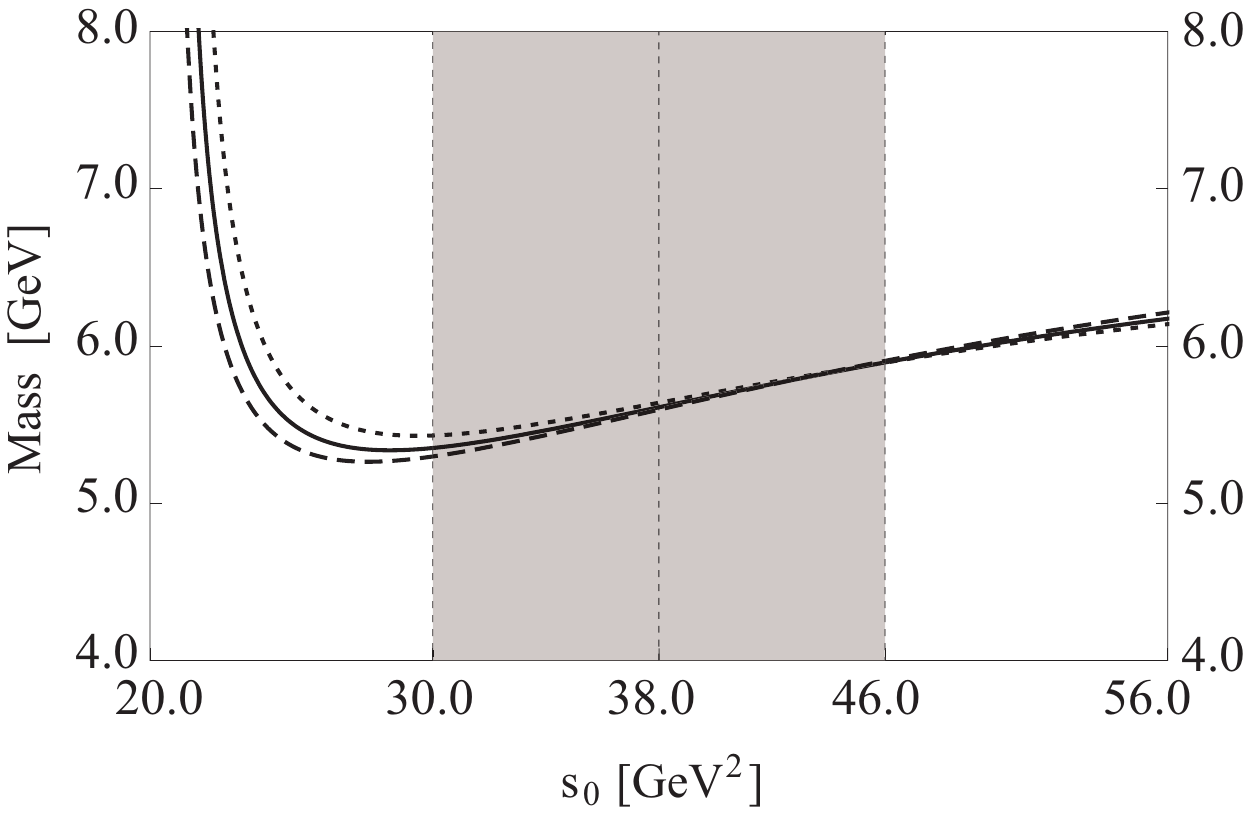}
~~~~~~~~~~
\includegraphics[width=0.4\textwidth]{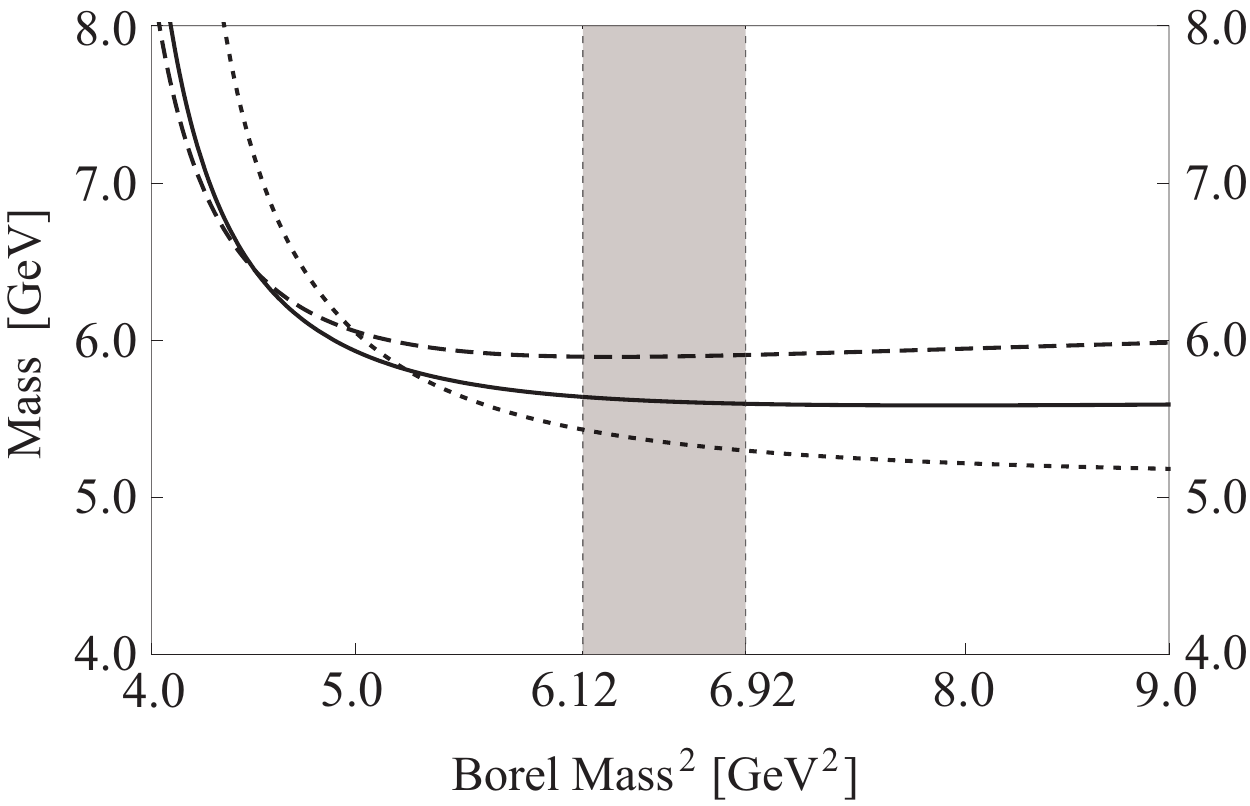}
\caption{Mass of the double-gluon hybrid state $|\bar q q gg;0^{++}\rangle$ as a function of the threshold value $s_0$ (left) and the Borel mass $M_B$ (right). In the left panel the dotted/solid/dashed curves are obtained by setting $M_B^2 = 6.12/6.52/6.92$ GeV$^2$, respectively. In the right panel the dotted/solid/dashed curves are obtained by setting $s_0 = 30.0/38.0/46.0$ GeV$^2$, respectively. These curves are obtained using the spectral density $\rho^{\bar q q gg}_{0^{++}}(s)$ extracted from the current $J_{0^{++}}$ with the quark-gluon content $\bar q q gg$ ($q=u/d$).}
\label{fig:mass}
\end{center}
\end{figure*}

\begin{figure*}[hbtp]
\begin{center}
\includegraphics[width=0.4\textwidth]{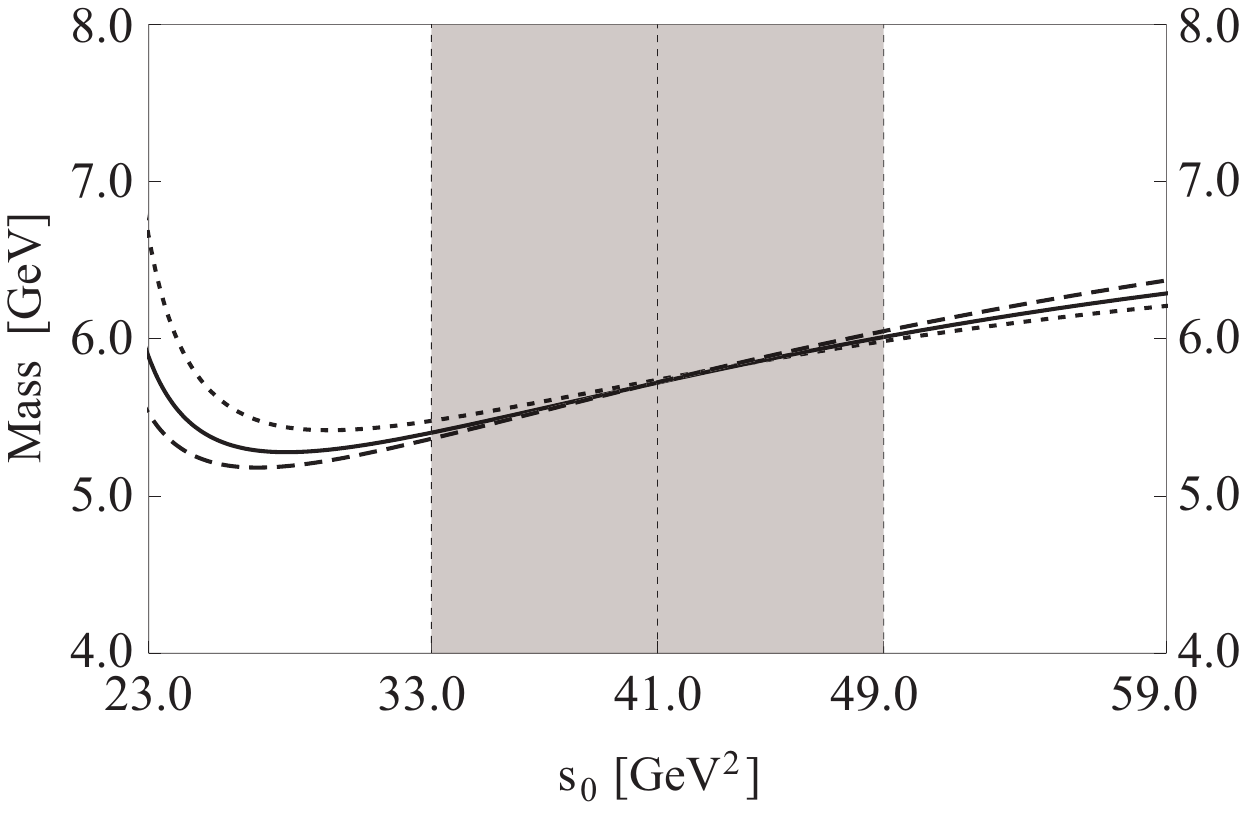}
~~~~~~~~~~
\includegraphics[width=0.4\textwidth]{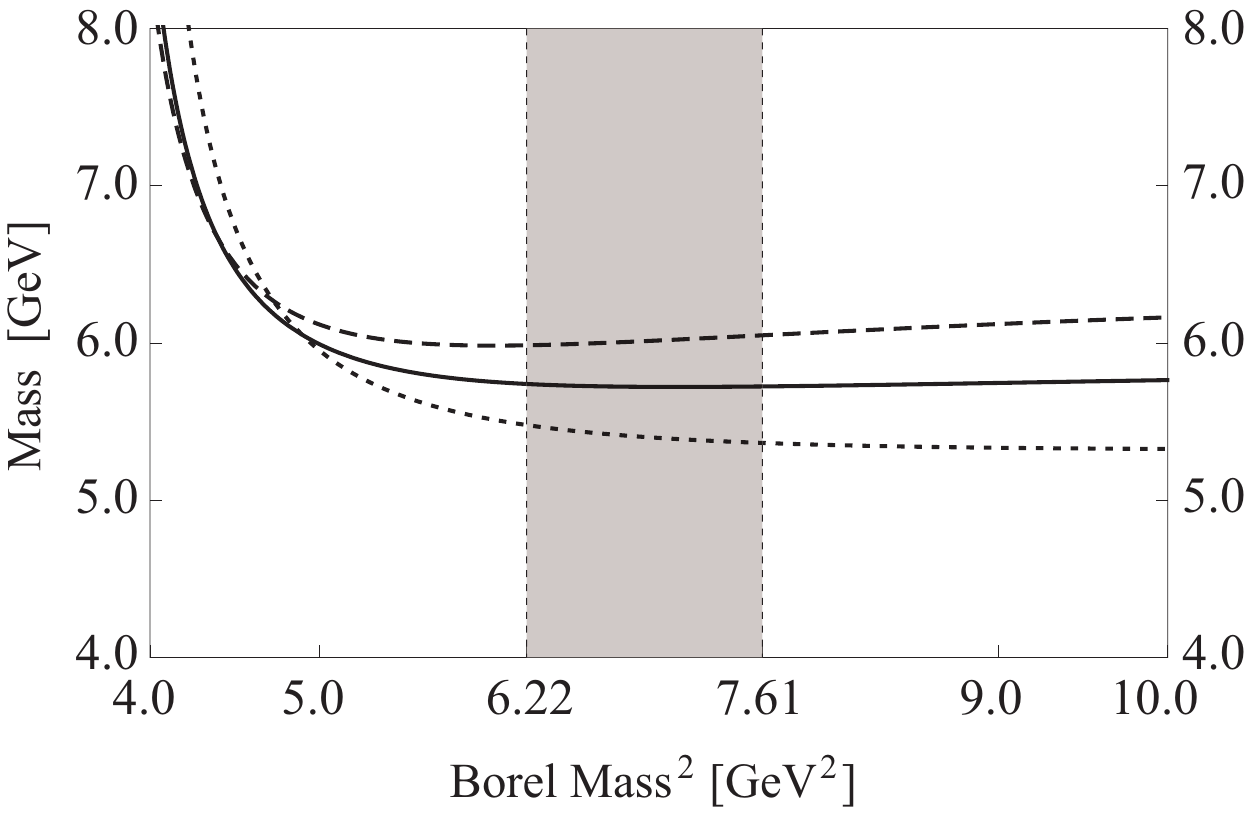}
\caption{Mass of the double-gluon hybrid state $|\bar s s gg;0^{++}\rangle$ as a function of the threshold value $s_0$ (left) and the Borel mass $M_B$ (right). In the left panel the dotted/solid/dashed curves are obtained by setting $M_B^2 = 6.22/6.91/7.61$ GeV$^2$, respectively. In the right panel the dotted/solid/dashed curves are obtained by setting $s_0 = 33.0/41.0/49.0$ GeV$^2$, respectively. These curves are obtained using the spectral density $\rho^{\bar s s gg}_{0^{++}}(s)$ extracted from the current $J_{0^{++}}$ with the quark-gluon content $\bar s s gg$.}
\label{fig:massssgg}
\end{center}
\end{figure*}

Altogether we determine the Borel window to be $6.12$~GeV$^2 \leq M_B^2 \leq 6.92$~GeV$^2$ for $s_0 = 38.0$~GeV$^2$. We redo the same procedures by changing $s_0$, and find that there are non-vanishing Borel windows as long as $s_0 \geq s^{\rm min}_0 = 34.9$~GeV$^2$. We choose $s_0$ to be slightly larger, and determine the working regions to be $30.0$~GeV$^2 \leq s_0 \leq 46.0$~GeV$^2$ and $6.12$~GeV$^2 \leq M_B^2 \leq 6.92$~GeV$^2$, where we calculate the mass of the double-gluon hybrid state $|\bar q q gg;0^{++}\rangle$ to be
\begin{equation}
M_{|\bar q q gg;0^{++}\rangle} = 5.61^{+0.29}_{-0.27}{\rm~GeV} \, .
\end{equation}
Its uncertainty is due to the threshold value $s_0$, the Borel mass $M_B$, and the QCD parameters listed in Eqs.~(\ref{eq:condensate}).

We show the mass $M_{|\bar q q gg;0^{++}\rangle}$ in Fig.~\ref{fig:mass} as a function of the threshold value $s_0$ and the Borel mass $M_B$. From the left panel, we find a mass minimum around $s_0 \sim 28$~GeV$^2$, and the $s_0$ dependence is acceptable inside the region $30.0$~GeV$^2 \leq s_0 \leq 46.0$~GeV$^2$. From the right panel, we find that the mass curves are sufficiently stable inside the region $6.12$~GeV$^2 \leq M_B^2 \leq 6.92$~GeV$^2$.

Similarly, we perform numerical analyses using the rest of the double-gluon hybrid currents with the quark-gluon content $\bar q q gg$ ($q=u/d$). The obtained results are summarized in Table~\ref{tab:results}. Besides, we also perform numerical analyses using these currents with the quark-gluon content $\bar s s gg$. The obtained results are also summarized in Table~\ref{tab:results}. Especially, the mass of the double-gluon hybrid state $|\bar s s gg;0^{++}\rangle$ is calculated to be
\begin{equation}
M_{|\bar s s gg;0^{++}\rangle} = 5.72^{+0.29}_{-0.32}{\rm~GeV} \, .
\end{equation}
For completeness, we show it in Fig.~\ref{fig:massssgg} as a function of the threshold value $s_0$ and the Borel mass $M_B$.

%
\section{Summary and Discussions}
\label{sec:summary}

\begin{table*}[hbtp]
\begin{center}
\renewcommand{\arraystretch}{1.6}
\caption{QCD sum rule results of the double-gluon hybrid states $|\bar q q gg; J^{PC}\rangle$ and $|\bar s s gg; J^{PC}\rangle$, extracted from the seven currents $J^{\cdots}_{0^{\pm+}/1^{\pm-}/2^{\pm+}/2^{+-}}$ with the quark-gluon contents $\bar q q gg$ ($q=u/d$) and $\bar s s gg$, respectively. The five currents $J^{\cdots}_{0^{\pm-}/1^{\pm+}/2^{--}}$ vanish, so their results are not given.}
\begin{tabular}{c|c|c|c|c|c|c}
\hline\hline
~~~\multirow{2}{*}{State [$J^{PC}$]}~~~ & ~~~~\multirow{2}{*}{Current}~~~~ & ~\multirow{2}{*}{~$s_0^{min}~[{\rm GeV}^2]$~}~ & \multicolumn{2}{c|}{Working Regions} & ~~\multirow{2}{*}{Pole~[\%]}~~ & ~\multirow{2}{*}{~Mass~[GeV]~}~
\\ \cline{4-5}
&  &  & ~~$M_B^2~[{\rm GeV}^2]$~~ & ~~$s_0~[{\rm GeV}^2]$~~ &&
\\ \hline\hline
$|\bar q q gg; 0^{++}\rangle$ & $J_{0^{++}}$                                         &  34.9   &  $6.12$--$6.92$   &  $38\pm8.0$  &  $40$--$50$  &  $5.61^{+0.29}_{-0.27}$
\\
$|\bar q q gg; 0^{-+}\rangle$ & $J_{0^{-+}}$                                         &  24.4   &  $5.34$--$5.78$   &  $27\pm5.0$  &  $40$--$48$  &  $4.25^{+0.32}_{-0.39}$
\\
$|\bar q q gg; 1^{+-}\rangle$ & $J^{\alpha\beta}_{1^{+-}}$                           &  32.1   &  $5.51$--$6.31$   &  $35\pm7.0$  &  $40$--$50$  &  $5.46^{+0.25}_{-0.18}$
\\
$|\bar q q gg; 1^{--}\rangle$ & $J^{\alpha\beta}_{1^{--}}$                           &  20.0   &  $4.60$--$4.91$   &  $22\pm4.0$  &  $40$--$47$  &  $3.74^{+0.30}_{-0.35}$
\\
$|\bar q q gg; 2^{++}\rangle$ & $J^{\alpha_1\beta_1,\alpha_2\beta_2}_{2^{++}}$       &  20.0   &  $5.39$--$5.76$   &  $22\pm4.0$  &  $40$--$46$  &  $3.74^{+0.27}_{-0.32}$
\\
$|\bar q q gg; 2^{+-}\rangle$ & $J^{\alpha_1\beta_1,\alpha_2\beta_2}_{2^{+-}}$       &  6.4    &  $1.61$--$1.78$   &  $7\pm2.0$   &  $40$--$48$  &  $2.26^{+0.20}_{-0.25}$
\\
$|\bar q q gg; 2^{-+}\rangle$ & $J^{\alpha_1\beta_1,\alpha_2\beta_2}_{2^{-+}}$       &  16.8   &  $4.39$--$4.81$   &  $19\pm4.0$  &  $40$--$49$  &  $3.51^{+0.29}_{-0.35}$
\\ \hline\hline
$|\bar s s gg; 0^{++}\rangle$ & $J_{0^{++}}$                                         &  35.3   &  $6.22$--$7.61$   &  $41\pm8.0$  &  $40$--$57$  &  $5.72^{+0.29}_{-0.32}$
\\
$|\bar s s gg; 0^{-+}\rangle$ & $J_{0^{-+}}$                                         &  24.5   &  $5.36$--$5.95$   &  $28\pm6.0$  &  $40$--$50$  &  $4.34^{+0.36}_{-0.46}$
\\
$|\bar s s gg; 1^{+-}\rangle$ & $J^{\alpha\beta}_{1^{+-}}$                           &  32.5   &  $5.60$--$6.79$   &  $37\pm8.0$  &  $40$--$55$  &  $5.52^{+0.29}_{-0.27}$
\\
$|\bar s s gg; 1^{--}\rangle$ & $J^{\alpha\beta}_{1^{--}}$                           &  20.2   &  $4.62$--$5.07$   &  $23\pm5.0$  &  $40$--$50$  &  $3.84^{+0.35}_{-0.44}$
\\
$|\bar s s gg; 2^{++}\rangle$ & $J^{\alpha_1\beta_1,\alpha_2\beta_2}_{2^{++}}$       &  20.4   &  $5.45$--$6.11$   &  $24\pm5.0$  &  $40$--$51$  &  $3.91^{+0.32}_{-0.39}$
\\
$|\bar s s gg; 2^{+-}\rangle$ & $J^{\alpha_1\beta_1,\alpha_2\beta_2}_{2^{+-}}$       &  7.1    &  $1.79$--$2.01$   & $8\pm2.0$    &  $40$--$50$  &  $2.38^{+0.19}_{-0.25}$
\\
$|\bar s s gg; 2^{-+}\rangle$ & $J^{\alpha_1\beta_1,\alpha_2\beta_2}_{2^{-+}}$       &  17.1   &  $4.44$--$5.00$   &  $20\pm4.0$  &  $40$--$51$  &  $3.61^{+0.28}_{-0.34}$
\\ \hline\hline
\end{tabular}
\label{tab:results}
\end{center}
\end{table*}

In this paper we study the double-gluon hybrid states with the quark-gluon contents $\bar q q gg$ ($q=u/d$) and $\bar s s gg$. We systematically construct twelve double-gluon hybrid currents using the color-octet quark-antiquark field
\begin{equation}
\nonumber \bar q_a \gamma_5 \lambda_n^{ab} q_b \, ,
\end{equation}
and the color-octet double-gluon fields
\begin{equation}
\nonumber d^{npq} G_p^{\alpha\beta} G_q^{\gamma\delta} \, , \, f^{npq} G_p^{\alpha\beta} G_q^{\gamma\delta} \, .
\end{equation}
Since the field $\bar q_a \gamma_5 \lambda_n^{ab} q_b$ has the $S$-wave spin-parity quantum number $J^P = 0^-$, these currents may couple to the lowest-lying double-gluon hybrid states.

We apply the method of QCD sum rules to study these currents, and the obtained results are summarized in Table~\ref{tab:results}. Note that the five currents $J^{\cdots}_{0^{\pm-}/1^{\pm+}/2^{--}}$ vanish due to some internal symmetries between the two gluon fields, but this does not indicate that the double-gluon hybrid states with these quantum numbers do not exist, since more currents can be constructed by combining the color-octet quark-antiquark fields
\begin{equation}
\nonumber \bar q_a \lambda_n^{ab} q_b \, , \,  \bar q_a \lambda_n^{ab} \gamma_\mu q_b \, , \, \bar q_a \lambda_n^{ab} \gamma_\mu \gamma_5 q_b \, , \, \bar q_a \lambda_n^{ab} \sigma_{\mu\nu} q_b \, ,
\end{equation}
and the above color-octet double-gluon fields. We shall systematically investigate them in the near future.

As shown in Table~\ref{tab:results}, only the masses of the double-gluon hybrid states $|\bar q q gg;2^{+-}\rangle$ and $|\bar s s gg;2^{+-}\rangle$ are calculated to be smaller than 3.0~GeV:
\begin{eqnarray}
\nonumber M_{|\bar q q gg;2^{+-}\rangle} &=& 2.26^{+0.20}_{-0.25}{\rm~GeV} \, ,
\\ \nonumber M_{|\bar s s gg;2^{+-}\rangle} &=& 2.38^{+0.19}_{-0.25}{\rm~GeV} \, .
\end{eqnarray}
These two states are coupled by the current $J^{\alpha_1\beta_1,\alpha_2\beta_2}_{2^{+-}}$ with the quark-gluon contents $\bar q q gg$ ($q=u/d$) and $\bar s s gg$. They are quite interesting because they both have the exotic quantum number $J^{PC} = 2^{+-}$ that can not be reached by the conventional $\bar q q$ mesons. Since we do not differentiate the up and down quarks within the QCD sum rule framework, the masses of the double-gluon hybrid states in the same isospin multiplet are calculated to be the same:
\begin{eqnarray}
\nonumber M_{|\bar q q gg;1^+2^{+-}\rangle} = M_{|\bar q q gg;0^-2^{+-}\rangle} &=& 2.26^{+0.20}_{-0.25}{\rm~GeV} \, ,
\\ \nonumber M_{|\bar s s gg;0^-2^{+-}\rangle} &=& 2.38^{+0.19}_{-0.25}{\rm~GeV} \, .
\end{eqnarray}

\begin{figure}[hbtp]
\begin{center}
\scalebox{0.24}{\includegraphics{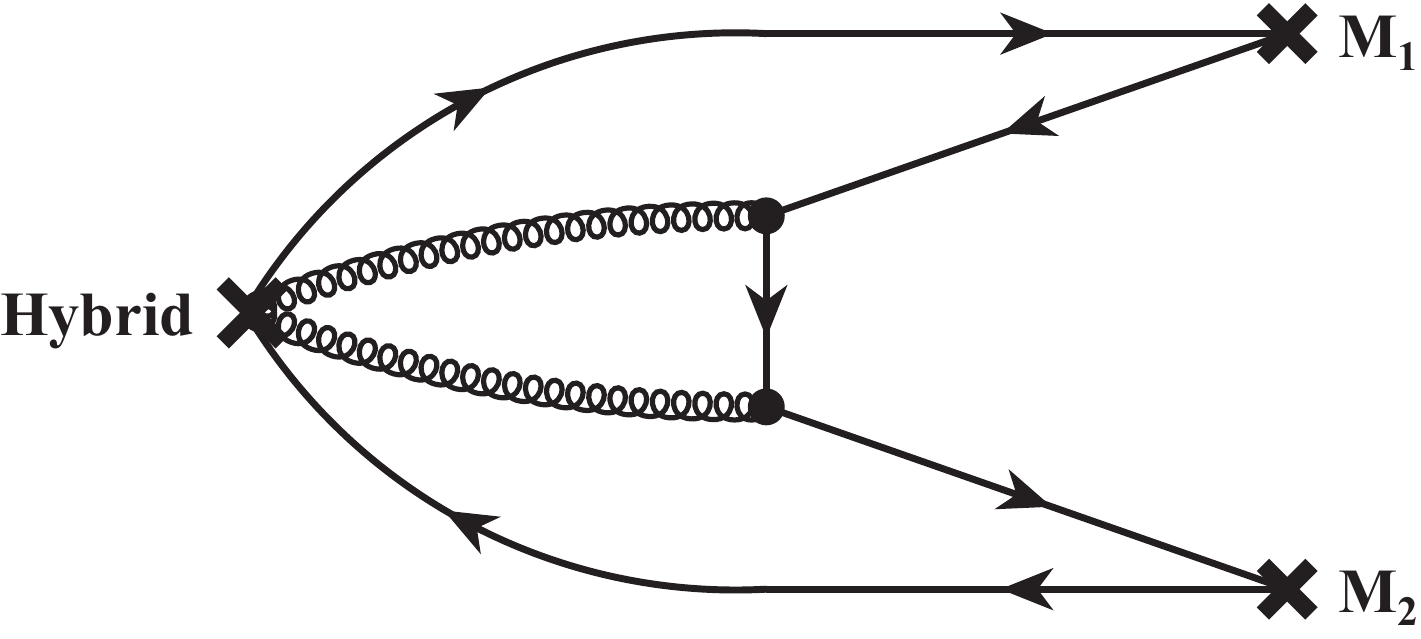}}
\scalebox{0.24}{\includegraphics{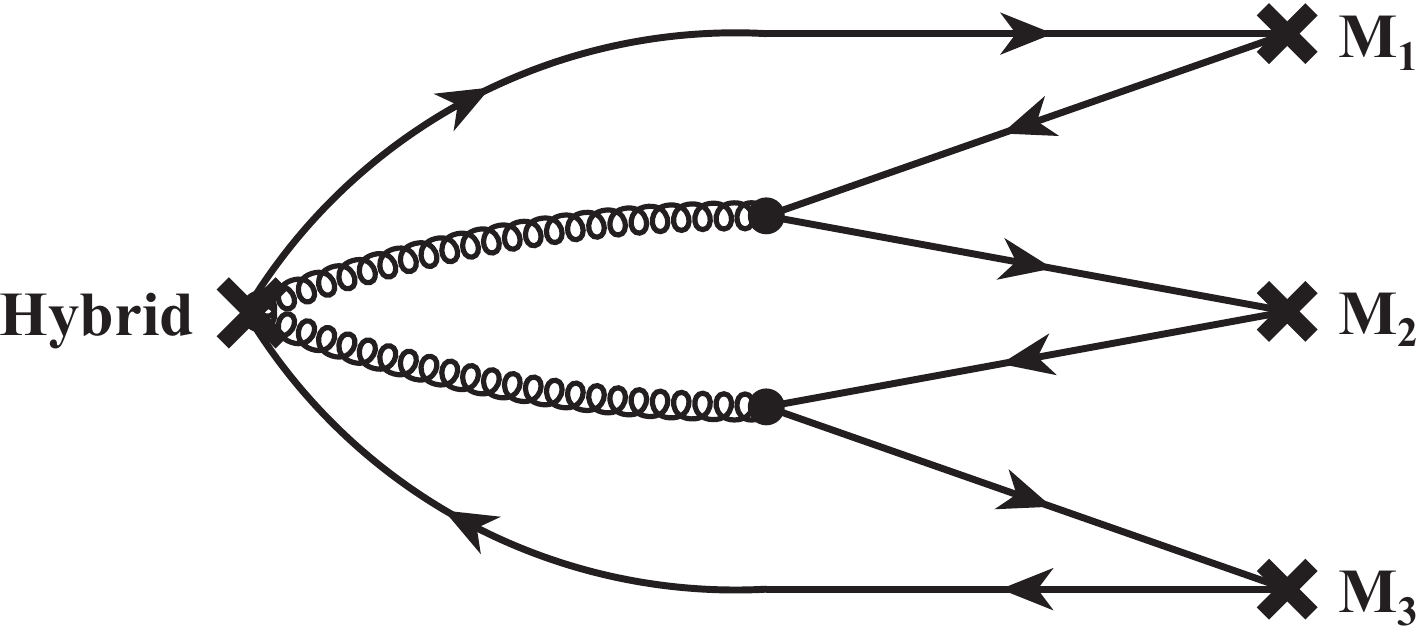}}
\end{center}
\caption{Two- and three-meson decay processes of the double-gluon hybrid state.}
\label{fig:decay}
\end{figure}

\begin{table*}[]
\begin{center}
\renewcommand{\arraystretch}{1.5}
\caption{Possible two- and three-meson decay patterns of the double-gluon hybrid states $|\bar q q gg;1^+2^{+-}\rangle$, $|\bar q q gg;0^-2^{+-}\rangle$, and $|\bar s s gg;0^-2^{+-}\rangle$. The results of the former two are partly taken from Ref.~\cite{Chen:2021smz}. We have used the notations: $h_1 = h_1(1170)$, $h_1^\prime = h_1(1415)$, $f_1 = f_1(1285)$, $f_1^\prime = f_1(1420)$, $a_0 = a_0(980)$, $f_0 = f_0(980)$, $a_1 = a_1(1260)$, $b_1 = b_1(1235)$, $a_2 = a_2(1320)$, $K_0^* = K_0^*(700)$, $K_1 = K_1(1270)/K_1(1400)$, and $K_2^* = K_2^*(1430)$.}
\begin{tabular}{ c | c | c | c }
\hline\hline
Two-Meson               & ~~~~~~~~~~~$|\bar q q gg;1^+2^{+-}\rangle$~~~~~~~~~~~                                 & ~~~~~~~~$|\bar q q gg;0^-2^{+-}\rangle$~~~~~~~~               & ~~~~~~~~$|\bar s s gg;0^-2^{+-}\rangle$~~~~~~~~
\\ \hline\hline
\multirow{1}{*}{$S$-wave} & \multicolumn{3}{c}{$K_2^*\bar K_0^*$}
\\ \hline
\multirow{2}{*}{$P$-wave} & ~~~$h_1\pi, a_1\pi, a_2\pi, b_1 \eta, b_1 \eta^\prime, \rho f_0,\omega a_0$~~~      & $b_1 \pi, h_1\eta, h_1\eta^\prime, \rho a_0,\omega f_0$                  & $h_1^\prime \eta, h_1^\prime \eta^\prime$
\\ \cline{2-4}
                        & \multicolumn{3}{c}{$K_1 \bar K, K_1 \bar K^*, K_2^* \bar K, K^* \bar K_0^*$}
\\ \hline
\multirow{2}{*}{$D$-wave} & $\rho^+\rho^-, \omega \pi, \rho\eta, \rho\eta^\prime$                               & $\rho \pi, \omega \eta, \omega \eta^\prime$                   & $\phi \eta, \phi \eta^\prime$
\\ \cline{2-4}
                        & \multicolumn{3}{c}{$K^* \bar K, K^* \bar K^*, K_1 \bar K_0^*$}
\\ \hline\hline
~Three-Meson~           & ~~~~~~~~~~~$|\bar q q gg;1^+2^{+-}\rangle$~~~~~~~~~~~                                 & ~~~~~~~~$|\bar q q gg;0^-2^{+-}\rangle$~~~~~~~~               & ~~~~~~~~$|\bar s s gg;0^-2^{+-}\rangle$~~~~~~~~
\\ \hline\hline
$S$-wave                  & $f_1\omega\pi, a_1\rho\pi$                                                          & $f_1\rho\pi, a_1\omega\pi$                                    & --
\\ \hline
\multirow{2}{*}{$P$-wave} & $\rho \pi \pi, \omega \eta \pi, \omega \eta^\prime \pi, \rho \eta \eta$             & $\omega \pi \pi, \rho \eta \pi, \rho \eta^\prime \pi,\omega \eta \eta$         & $\phi \eta \eta$
\\ \cline{2-4}
                        & \multicolumn{3}{c}{$\pi K^* \bar K, \rho K \bar K, \omega K \bar K, \phi K \bar K, \pi K^* \bar K^*, \rho K^* \bar K, \omega K^* \bar K, \eta K^* \bar K$}
\\ \hline\hline
\end{tabular}
\label{tab:decay}
\end{center}
\end{table*}

The double-gluon hybrid states ${|\bar q q gg;1^+2^{+-}\rangle}$ and ${|\bar q q gg;0^-2^{+-}\rangle}$ have been studied in Ref.~\cite{Chen:2021smz}, and their possible decay patterns have also been partly derived there. In this paper we further study the decay patterns of its partner state $|\bar s s gg;0^-2^{+-}\rangle$. As shown in Fig.~\ref{fig:decay}, a double-gluon hybrid state can decay after exciting two $\bar q q/\bar s s$ pairs from two gluons, followed by reorganizing three color-octet $\bar q q/\bar s s$ pairs into two or three color-singlet mesons. The amplitudes of these two decay processes are both at the $\mathcal{O}(\alpha_s)$ order, so the three-meson decay patterns are not suppressed severely compared to the two-meson decay patterns.

We list in Table~\ref{tab:decay} some possible two- and three-meson decay patterns for the double-gluon hybrid states $|\bar q q gg;1^+2^{+-}\rangle$, $|\bar q q gg;0^-2^{+-}\rangle$, and $|\bar s s gg;0^-2^{+-}\rangle$. Accordingly, we propose to search for $|\bar q q gg;1^+2^{+-}\rangle$ in the two-meson decay channels $\rho f_0(980)/\omega \pi/K^* \bar K$ and the three-meson decay channels $f_1\omega\pi/\rho\pi\pi$, etc.; we propose to search for $|\bar q q gg;0^-2^{+-}\rangle$ in the two-meson decay channels $\rho a_0(980)/\rho\pi/K^* \bar K$ and the three-meson decay channels $f_1\rho\pi/\omega\pi\pi$, etc.; we propose to search for $|\bar s s gg;0^-2^{+-}\rangle$ in the two-meson decay channels $\phi \eta/K^* \bar K$ and the three-meson decay channel $\rho K \bar K$, etc.

%
\section*{Acknowledgments}
%

This project is supported by
the National Natural Science Foundation of China under Grant No. 11722540, No.~11975033, No. 12075019, and No.~12070131001,
the National Key R$\&$D Program of China under Contracts No. 2020YFA0406400,
the Jiangsu Provincial Double-Innovation Program under Grant No.~JSSCRC2021488,
and
the Fundamental Research Funds for the Central Universities.

\appendix
\section{Spectral densities}
\label{app:sumrule}

In this appendix we list the spectral densities extracted from the double-gluon hybrid currents $J^{\cdots}_{0^{-+}/1^{\pm-}/2^{\pm+}/2^{+-}}$ with the quark-gluon contents $\bar q q gg$ ($q=u/d$) and $\bar s s gg$. The five currents $J^{\cdots}_{0^{\pm-}/1^{\pm+}/2^{--}}$ vanish, so their results are not given.
\begin{widetext}
\begin{eqnarray}
\nonumber \rho_{0^{-+}}^{\bar{q}qgg}(s) &=& \frac{\alpha_s^2 s^5}{4320 \pi^4} + \frac{35\alpha_s^2 \langle g_s^2 GG \rangle s^3}{3456 \pi^4} + \left( \frac{80 \alpha_s^2 \langle \bar q q \rangle^2}{27} + \frac{5 \alpha_s \langle g_s^3 G^3 \rangle}{144\pi^3} \right) s^2 + \left( - \frac{80\alpha_s^2 \langle \bar q q \rangle \langle \bar g_s q \sigma G q \rangle}{9} + \frac{5 \langle g_s^2 GG \rangle^2}{288\pi^2} \right) s \, ,
\\
\\
\nonumber \rho_{0^{-+}}^{\bar{s}sgg}(s) &=& \frac{\alpha_s^2 s^5}{4320 \pi^4}- \frac{\alpha_s^2 m_s^2 s^4}{144 \pi^4}  + \left(  \frac{35\alpha_s^2 \langle g_s^2 GG \rangle }{3456 \pi^4} - \frac{5\alpha_s^2 m_s \langle \bar s s \rangle }{27 \pi^2} + \frac{5\alpha_s^2 m_s^4 }{36 \pi^4}\right) s^3
\\ \nonumber &&+ \left( \frac{80 \alpha_s^2 \langle \bar s s \rangle^2}{27} + \frac{5 \alpha_s \langle g_s^3 G^3 \rangle}{144\pi^3}-\frac{20 \alpha_s^2 m_s^3  \langle \bar s s \rangle}{9\pi^2} +\frac{10 \alpha_s^2 m_s  \langle g_s \bar s \sigma G s \rangle}{9\pi^2}-\frac{25 \alpha_s^2 m_s^2  \langle g_s^2 GG \rangle}{192\pi^4} \right) s^2
\\ \nonumber && + \left( \frac{5 \langle g_s^2 GG \rangle^2}{288\pi^2}- \frac{80\alpha_s^2 \langle \bar s s \rangle \langle g_s \bar s \sigma G s \rangle}{9} -\frac{5 \alpha_s m_s^2 \langle g_s^3 G^3 \rangle}{24\pi^3} +\frac{40 \alpha_s^2 m_s^2 \langle \bar s s \rangle^2}{9} \right.
\\ &&
\left.-\frac{25 \alpha_s^2 m_s \langle g_s^2 GG \rangle \langle \bar s s \rangle}{24\pi^2}+\frac{25 \alpha_s^2 m_s^4 \langle g_s^2 GG \rangle }{32\pi^4} \right) s \, ,
\\
\rho_{1^{+-}}^{\bar{q}qgg}(s) &=& \frac{\alpha_s^2 s^5}{20160 \pi^4} - \frac{7\alpha_s^2 \langle g_s^2 GG \rangle s^3}{15360 \pi^4} + \left( \frac{4 \alpha_s^2 \langle \bar q q \rangle^2}{9} - \frac{ \alpha_s \langle g_s^3 G^3 \rangle}{64\pi^3} \right) s^2 + \left( - \frac{8\alpha_s^2 \langle \bar q q \rangle \langle \bar g_s q \sigma G q \rangle}{9} - \frac{ \langle g_s^2 GG \rangle^2}{192\pi^2} \right) s \, ,
\\
\nonumber \rho_{1^{+-}}^{\bar{s}sgg}(s) &=& \frac{\alpha_s^2 s^5}{20160 \pi^4}- \frac{\alpha_s^2 m_s^2 s^4}{720 \pi^4}  + \left(  -\frac{7\alpha_s^2 \langle g_s^2 GG \rangle }{15360 \pi^4} - \frac{\alpha_s^2 m_s \langle \bar s s \rangle }{30 \pi^2} + \frac{\alpha_s^2 m_s^4 }{40 \pi^4}\right) s^3
\\ \nonumber &&+ \left( \frac{4 \alpha_s^2 \langle \bar s s \rangle^2}{9} - \frac{ \alpha_s \langle g_s^3 G^3 \rangle}{64\pi^3}-\frac{\alpha_s^2 m_s^3  \langle \bar s s \rangle}{3\pi^2} +\frac{ \alpha_s^2 m_s  \langle g_s \bar s \sigma G s \rangle}{6\pi^2}+\frac{5 \alpha_s^2 m_s^2  \langle g_s^2 GG \rangle}{1024\pi^4} \right) s^2
\\ \nonumber &&+ \left(- \frac{ \langle g_s^2 GG \rangle^2}{192\pi^2}- \frac{8\alpha_s^2 \langle \bar s s \rangle \langle g_s \bar s \sigma G s \rangle}{9} +\frac{ \alpha_s m_s^2 \langle g_s^3 G^3 \rangle}{12\pi^3}+\frac{4 \alpha_s^2 m_s^2 \langle \bar s s \rangle^2}{9}\right.
\\ &&
\left.+\frac{5 \alpha_s^2 m_s \langle g_s^2 GG \rangle \langle \bar s s \rangle}{96\pi^2}-\frac{5 \alpha_s^2 m_s^4 \langle g_s^2 GG \rangle }{128\pi^4} \right) s \, ,
\\
\rho_{1^{--}}^{\bar{q}qgg}(s) &=& \frac{\alpha_s^2 s^5}{20160 \pi^4} + \frac{3\alpha_s^2 \langle g_s^2 GG \rangle s^3}{2560 \pi^4} + \left( \frac{4 \alpha_s^2 \langle \bar q q \rangle^2}{9} + \frac{ \alpha_s \langle g_s^3 G^3 \rangle}{192\pi^3} \right) s^2 + \left( - \frac{8\alpha_s^2 \langle \bar q q \rangle \langle \bar g_s q \sigma G q \rangle}{9} + \frac{ \langle g_s^2 GG \rangle^2}{192\pi^2} \right) s \, ,
\\
\nonumber \rho_{1^{--}}^{\bar{s}sgg}(s) &=& \frac{\alpha_s^2 s^5}{20160 \pi^4}- \frac{\alpha_s^2 m_s^2 s^4}{720 \pi^4}  + \left(  \frac{3\alpha_s^2 \langle g_s^2 GG \rangle }{2560 \pi^4} - \frac{\alpha_s^2 m_s \langle \bar s s \rangle }{30 \pi^2} + \frac{\alpha_s^2 m_s^4 }{40 \pi^4}\right) s^3
\\ \nonumber &&+ \left( \frac{4 \alpha_s^2 \langle \bar s s \rangle^2}{9} + \frac{ \alpha_s \langle g_s^3 G^3 \rangle}{192\pi^3}-\frac{ \alpha_s^2 m_s^3  \langle \bar s s \rangle}{3\pi^2} +\frac{ \alpha_s^2 m_s  \langle g_s \bar s \sigma G s \rangle}{6\pi^2}-\frac{15 \alpha_s^2 m_s^2  \langle g_s^2 GG \rangle}{1024\pi^4} \right) s^2
\\ \nonumber &&+ \left( \frac{ \langle g_s^2 GG \rangle^2}{192\pi^2}- \frac{8\alpha_s^2 \langle \bar s s \rangle \langle g_s \bar s \sigma G s \rangle}{9} -\frac{ \alpha_s m_s^2 \langle g_s^3 G^3 \rangle}{24\pi^3}+\frac{4 \alpha_s^2 m_s^2 \langle \bar s s \rangle^2}{9}\right.
\\ &&
\left.-\frac{5 \alpha_s^2 m_s \langle g_s^2 GG \rangle \langle \bar s s \rangle}{48\pi^2}+\frac{5 \alpha_s^2 m_s^4 \langle g_s^2 GG \rangle }{64\pi^4} \right) s \, ,
\\ \nonumber
\rho_{2^{++}}^{\bar{q}qgg}(s) &=& \frac{\alpha_s^2 s^5}{483840 \pi^4} + \left( \frac{\alpha_s \langle g_s^2 GG \rangle}{6912 \pi^3} + \frac{19\alpha_s^2 \langle g_s^2 GG \rangle}{248832 \pi^4} \right) s^3 + \left( \frac{\alpha_s^2 \langle \bar q q \rangle^2}{27} - \frac{\alpha_s \langle g_s^3 G^3 \rangle}{13824\pi^3} \right) s^2
\\ && + \left( - \frac{10\alpha_s^2 \langle \bar q q \rangle \langle \bar g_s q \sigma G q \rangle}{81} - \frac{5\alpha_s \langle g_s^2 GG \rangle^2}{165888\pi^3}+\frac{5 \langle g_s^2 GG \rangle^2}{6912\pi^2} \right) s \, ,
\\
\nonumber \rho_{2^{++}}^{\bar{s}sgg}(s) &=& \frac{\alpha_s^2 s^5}{483840 \pi^4}- \frac{\alpha_s^2 m_s^2 s^4}{72576 \pi^4}  + \left(  \frac{19\alpha_s^2 \langle g_s^2 GG \rangle }{248832 \pi^4} - \frac{\alpha_s^2 m_s \langle \bar s s \rangle }{972 \pi^2} + \frac{\alpha_s^2 m_s^4 }{648 \pi^4}+\frac{\alpha_s \langle g_s^2 GG \rangle }{6912 \pi^3} \right) s^3
\\ \nonumber &&+ \left( \frac{ \alpha_s^2 \langle \bar s s \rangle^2}{27} - \frac{ \alpha_s \langle g_s^3 G^3 \rangle}{13824\pi^3}-\frac{ \alpha_s^2 m_s^3  \langle \bar s s \rangle}{36\pi^2} +\frac{ \alpha_s^2 m_s  \langle g_s \bar s \sigma G s \rangle}{72\pi^2}-\frac{ \alpha_s^2 m_s^2  \langle g_s^2 GG \rangle}{1152\pi^4}-\frac{ 5\alpha_s m_s^2  \langle g_s^2 GG \rangle}{3456\pi^3} \right) s^2
\\ \nonumber &&+ \left( \frac{ 5\langle g_s^2 GG \rangle^2}{6912\pi^2}- \frac{35\alpha_s^2 m_s\langle \bar s s \rangle \langle g_s^2 GG \rangle}{6912\pi^2} +\frac{ 5\alpha_s m_s^2 \langle g_s^3 G^3 \rangle}{3456\pi^3}+\frac{35 \alpha_s^2 m_s^4 \langle g_s^2 GG \rangle}{9216\pi^4}+\frac{5 \alpha_s^2 m_s^2 \langle \bar s s \rangle^2}{81}\right.
\\ &&
\left.-\frac{5 \alpha_s m_s \langle g_s^2 GG \rangle \langle \bar s s \rangle}{648\pi}+\frac{5 \alpha_s m_s^4 \langle g_s^2 GG \rangle }{864\pi^3}-\frac{ 5\alpha_s \langle g_s^2 GG \rangle^2}{165888\pi^3} - \frac{10\alpha_s^2 \langle \bar q q \rangle \langle \bar g_s q \sigma G q \rangle}{81}\right) s \, ,
\\ \nonumber
\rho_{2^{+-}}^{\bar{q}qgg}(s) &=& \frac{\alpha_s^2 s^5}{80640 \pi^4} + \left( \frac{\alpha_s \langle g_s^2 GG \rangle}{3840 \pi^3} + \frac{7\alpha_s^2 \langle g_s^2 GG \rangle}{61440 \pi^4} \right) s^3 + \left( \frac{\alpha_s^2 \langle \bar q q \rangle^2}{9} - \frac{\alpha_s \langle g_s^3 G^3 \rangle}{1536\pi^3} \right) s^2
\\ && + \left( - \frac{2\alpha_s^2 \langle \bar q q \rangle \langle \bar g_s q \sigma G q \rangle}{9} - \frac{\alpha_s \langle g_s^2 GG \rangle^2}{18432\pi^3} \right) s \, ,
\\
\nonumber \rho_{2^{+-}}^{\bar{s}sgg}(s) &=& \frac{\alpha_s^2 s^5}{80640 \pi^4}- \frac{\alpha_s^2 m_s^2 s^4}{2880 \pi^4}  + \left(  \frac{7\alpha_s^2 \langle g_s^2 GG \rangle }{61440 \pi^4} - \frac{\alpha_s^2 m_s \langle \bar s s \rangle }{120 \pi^2} + \frac{\alpha_s^2 m_s^4 }{160 \pi^4}+\frac{\alpha_s \langle g_s^2 GG \rangle }{3840 \pi^3} \right) s^3
\\ \nonumber &&+ \left( \frac{ \alpha_s^2 \langle \bar s s \rangle^2}{9} - \frac{ \alpha_s \langle g_s^3 G^3 \rangle}{1536\pi^3}- \frac{ \alpha_s m_s^2 \langle g_s^2 GG \rangle}{384\pi^3}-\frac{ \alpha_s^2 m_s^3  \langle \bar s s \rangle}{12\pi^2} +\frac{ \alpha_s^2 m_s  \langle g_s \bar s \sigma G s \rangle}{24\pi^2}-\frac{3 \alpha_s^2 m_s^2  \langle g_s^2 GG \rangle}{2048\pi^4} \right) s^2
\\ \nonumber &&+ \left( -\frac{ \alpha_s\langle g_s^2 GG \rangle^2}{18432\pi^3}- \frac{\alpha_s^2 m_s\langle \bar s s \rangle \langle g_s^2 GG \rangle}{128\pi^2}- \frac{\alpha_s m_s\langle \bar s s \rangle \langle g_s^2 GG \rangle}{72\pi} +\frac{ \alpha_s m_s^2 \langle g_s^3 G^3 \rangle}{384\pi^3}+\frac{ \alpha_s m_s^4 \langle g_s^2 GG \rangle}{96\pi^3}\right.
\\ &&
\left.+\frac{3 \alpha_s^2 m_s^4 \langle g_s^2 GG \rangle }{512\pi^4}+\frac{ \alpha_s^2 m_s^2 \langle \bar s s \rangle^2 }{9}-\frac{ 2\alpha_s^2 \langle \bar s s \rangle\langle g_s \bar s \sigma G s \rangle}{9} \right) s \, ,
\\ \nonumber
\rho_{2^{-+}}^{\bar{q}qgg}(s) &=& \frac{\alpha_s^2 s^5}{414720 \pi^4} + \left( \frac{\alpha_s \langle g_s^2 GG \rangle}{6912 \pi^3} - \frac{25\alpha_s^2 \langle g_s^2 GG \rangle}{1990656 \pi^4} \right) s^3 + \left( \frac{\alpha_s^2 \langle \bar q q \rangle^2}{81} - \frac{7\alpha_s \langle g_s^3 G^3 \rangle}{13824\pi^3} \right) s^2
\\ && + \left( - \frac{5\alpha_s \langle g_s^2 GG \rangle^2}{165888\pi^3} +\frac{5 \langle g_s^2 GG \rangle^2}{13824\pi^3}\right) s \, ,
\\
\nonumber \rho_{2^{-+}}^{\bar{s}sgg}(s) &=& \frac{\alpha_s^2 s^5}{414720 \pi^4}- \frac{\alpha_s^2 m_s^2 s^4}{16128 \pi^4}  + \left(  -\frac{25\alpha_s^2 \langle g_s^2 GG \rangle }{1990656 \pi^4} - \frac{5\alpha_s^2 m_s \langle \bar s s \rangle }{3888 \pi^2} + \frac{5\alpha_s^2 m_s^4 }{5184 \pi^4}+\frac{\alpha_s \langle g_s^2 GG \rangle }{6912 \pi^3} \right) s^3
\\ \nonumber &&+ \left( \frac{ \alpha_s^2 \langle \bar s s \rangle^2}{81} - \frac{ 7\alpha_s \langle g_s^3 G^3 \rangle}{13824\pi^3}- \frac{ 5\alpha_s m_s^2 \langle g_s^2 GG \rangle}{3456\pi^3}-\frac{ \alpha_s^2 m_s^3  \langle \bar s s \rangle}{108\pi^2}
+\frac{ \alpha_s^2 m_s  \langle g_s \bar s \sigma G s \rangle}{216\pi^2}+\frac{5 \alpha_s^2 m_s^2  \langle g_s^2 GG \rangle}{36864\pi^4} \right) s^2
\\ \nonumber &&+ \left( -\frac{ 5\alpha_s\langle g_s^2 GG \rangle^2}{165888\pi^3} +\frac{ 5\langle g_s^2 GG \rangle^2}{13824\pi^2}+ \frac{25\alpha_s^2 m_s\langle \bar s s \rangle \langle g_s^2 GG \rangle}{13824\pi^2}+ \frac{5\alpha_s m_s^4 \langle g_s^2 GG \rangle}{864\pi^3} +\frac{ 5\alpha_s m_s^2 \langle g_s^3 G^3 \rangle}{3456\pi^3}\right.
\\ &&
\left.-\frac{ 25\alpha_s^2 m_s^4 \langle g_s^2 GG \rangle}{18432\pi^4}
-\frac{5 \alpha_s m_s \langle \bar s s \rangle\langle g_s^2 GG \rangle }{648\pi} \right) s \, .
\end{eqnarray}
\end{widetext}

\section*{References}
\bibliographystyle{elsarticle-num}
\bibliography{ref}

\begin{thebibliography}{10}
\expandafter\ifx\csname url\endcsname\relax
  \def\url#1{\texttt{#1}}\fi
\expandafter\ifx\csname urlprefix\endcsname\relax\def\urlprefix{URL }\fi
\expandafter\ifx\csname href\endcsname\relax
  \def\href#1#2{#2} \def\path#1{#1}\fi

\bibitem{pdg}
P.~A. Zyla, et~al., {Review of Particle Physics}, PTEP 2020~(8) (2020) 083C01.
\newblock \href {http://dx.doi.org/10.1093/ptep/ptaa104}
  {\path{doi:10.1093/ptep/ptaa104}}.

\bibitem{Chen:2022asf}
H.-X. Chen, W.~Chen, X.~Liu, Y.-R. Liu, S.-L. Zhu, {An updated review of the
  new hadron states}, Rept. Prog. Phys. 86~(2) (2023) 026201.
\newblock \href {http://arxiv.org/abs/2204.02649} {\path{arXiv:2204.02649}},
  \href {http://dx.doi.org/10.1088/1361-6633/aca3b6}
  {\path{doi:10.1088/1361-6633/aca3b6}}.

\bibitem{Klempt:2007cp}
E.~Klempt, A.~Zaitsev, {Glueballs, hybrids, multiquarks: Experimental facts
  versus QCD inspired concepts}, Phys. Rept. 454 (2007) 1--202.
\newblock \href {http://arxiv.org/abs/0708.4016} {\path{arXiv:0708.4016}},
  \href {http://dx.doi.org/10.1016/j.physrep.2007.07.006}
  {\path{doi:10.1016/j.physrep.2007.07.006}}.

\bibitem{Meyer:2015eta}
C.~A. Meyer, E.~S. Swanson, {Hybrid mesons}, Prog. Part. Nucl. Phys. 82 (2015)
  21--58.
\newblock \href {http://arxiv.org/abs/1502.07276} {\path{arXiv:1502.07276}},
  \href {http://dx.doi.org/10.1016/j.ppnp.2015.03.001}
  {\path{doi:10.1016/j.ppnp.2015.03.001}}.

\bibitem{Amsler:2004ps}
C.~Amsler, N.~A. Tornqvist, {Mesons beyond the naive quark model}, Phys. Rept.
  389 (2004) 61--117.
\newblock \href {http://dx.doi.org/10.1016/j.physrep.2003.09.003}
  {\path{doi:10.1016/j.physrep.2003.09.003}}.

\bibitem{Bugg:2004xu}
D.~V. Bugg, {Four sorts of meson}, Phys. Rept. 397 (2004) 257--358.
\newblock \href {http://arxiv.org/abs/hep-ex/0412045}
  {\path{arXiv:hep-ex/0412045}}, \href
  {http://dx.doi.org/10.1016/j.physrep.2004.03.008}
  {\path{doi:10.1016/j.physrep.2004.03.008}}.

\bibitem{Meyer:2010ku}
C.~A. Meyer, Y.~Van~Haarlem, {Status of exotic-quantum-number mesons}, Phys.
  Rev. C 82 (2010) 025208.
\newblock \href {http://arxiv.org/abs/1004.5516} {\path{arXiv:1004.5516}},
  \href {http://dx.doi.org/10.1103/PhysRevC.82.025208}
  {\path{doi:10.1103/PhysRevC.82.025208}}.

\bibitem{Briceno:2017max}
R.~A. Briceno, J.~J. Dudek, R.~D. Young, {Scattering processes and resonances
  from lattice QCD}, Rev. Mod. Phys. 90~(2) (2018) 025001.
\newblock \href {http://arxiv.org/abs/1706.06223} {\path{arXiv:1706.06223}},
  \href {http://dx.doi.org/10.1103/RevModPhys.90.025001}
  {\path{doi:10.1103/RevModPhys.90.025001}}.

\bibitem{COMPASS:2018uzl}
M.~Aghasyan, et~al., {Light isovector resonances in $\pi^- p \to
  \pi^-\pi^-\pi^+ p$ at 190 GeV/${\it c}$}, Phys. Rev. D 98~(9) (2018) 092003.
\newblock \href {http://arxiv.org/abs/1802.05913} {\path{arXiv:1802.05913}},
  \href {http://dx.doi.org/10.1103/PhysRevD.98.092003}
  {\path{doi:10.1103/PhysRevD.98.092003}}.

\bibitem{JPAC:2018zyd}
A.~Rodas, et~al., {Determination of the Pole Position of the Lightest Hybrid
  Meson Candidate}, Phys. Rev. Lett. 122~(4) (2019) 042002.
\newblock \href {http://arxiv.org/abs/1810.04171} {\path{arXiv:1810.04171}},
  \href {http://dx.doi.org/10.1103/PhysRevLett.122.042002}
  {\path{doi:10.1103/PhysRevLett.122.042002}}.

\bibitem{Ketzer:2019wmd}
B.~Ketzer, B.~Grube, D.~Ryabchikov, {Light-meson spectroscopy with COMPASS},
  Prog. Part. Nucl. Phys. 113 (2020) 103755.
\newblock \href {http://arxiv.org/abs/1909.06366} {\path{arXiv:1909.06366}},
  \href {http://dx.doi.org/10.1016/j.ppnp.2020.103755}
  {\path{doi:10.1016/j.ppnp.2020.103755}}.

\bibitem{Jin:2021vct}
S.~Jin, X.~Shen, {Highlights of light meson spectroscopy at the BESIII
  experiment}, Natl. Sci. Rev. 8~(11) (2021) nwab198.
\newblock \href {http://dx.doi.org/10.1093/nsr/nwab198}
  {\path{doi:10.1093/nsr/nwab198}}.

\bibitem{Meng:2022ozq}
L.~Meng, B.~Wang, G.-J. Wang, S.-L. Zhu, {Chiral perturbation theory for heavy
  hadrons and chiral effective field theory for heavy hadronic molecules}\href
  {http://arxiv.org/abs/2204.08716} {\path{arXiv:2204.08716}}.

\bibitem{IHEP-Brussels-LosAlamos-AnnecyLAPP:1988iqi}
D.~Alde, et~al., {Evidence for a $1^{-+}$ exotic meson}, Phys. Lett. B 205
  (1988) 397.
\newblock \href {http://dx.doi.org/10.1016/0370-2693(88)91686-3}
  {\path{doi:10.1016/0370-2693(88)91686-3}}.

\bibitem{E852:1998mbq}
G.~S. Adams, et~al., {Observation of a New $J^{PC}=1^{-+}$ Exotic State in the
  Reaction $\pi^- p \to \pi^+ \pi^- \pi^- p$ at 18~GeV$/c$}, Phys. Rev. Lett.
  81 (1998) 5760--5763.
\newblock \href {http://dx.doi.org/10.1103/PhysRevLett.81.5760}
  {\path{doi:10.1103/PhysRevLett.81.5760}}.

\bibitem{COMPASS:2009xrl}
M.~Alekseev, et~al., {Observation of a $J^{PC}=1^{-+}$ Exotic Resonance in
  Diffractive Dissociation of 190~GeV$/c$ $\pi^-$ into $\pi^- \pi^- \pi^+$},
  Phys. Rev. Lett. 104 (2010) 241803.
\newblock \href {http://arxiv.org/abs/0910.5842} {\path{arXiv:0910.5842}},
  \href {http://dx.doi.org/10.1103/PhysRevLett.104.241803}
  {\path{doi:10.1103/PhysRevLett.104.241803}}.

\bibitem{E852:2004gpn}
J.~Kuhn, et~al., {Exotic meson production in the $f_1(1285)\pi^-$ system
  observed in the reaction $\pi^- p \to \eta \pi^+ \pi^- \pi^- p$ at
  $18$~GeV$/c$}, Phys. Lett. B 595 (2004) 109--117.
\newblock \href {http://arxiv.org/abs/hep-ex/0401004}
  {\path{arXiv:hep-ex/0401004}}, \href
  {http://dx.doi.org/10.1016/j.physletb.2004.05.032}
  {\path{doi:10.1016/j.physletb.2004.05.032}}.

\bibitem{BESIII:2022riz}
M.~Ablikim, et~al., {Observation of an Isoscalar Resonance with Exotic
  $J^{PC}=1^{-+}$ Quantum Numbers in $J/\psi\rightarrow\gamma\eta\eta'$}, Phys.
  Rev. Lett. 129~(19) (2022) 192002.
\newblock \href {http://arxiv.org/abs/2202.00621} {\path{arXiv:2202.00621}},
  \href {http://dx.doi.org/10.1103/PhysRevLett.129.192002}
  {\path{doi:10.1103/PhysRevLett.129.192002}}.

\bibitem{BESIII:2022qzu}
M.~Ablikim, et~al., {Partial wave analysis of
  $J/\psi\rightarrow\gamma\eta\eta'$}, Phys. Rev. D 106~(7) (2022) 072012.
\newblock \href {http://arxiv.org/abs/2202.00623} {\path{arXiv:2202.00623}},
  \href {http://dx.doi.org/10.1103/PhysRevD.106.072012}
  {\path{doi:10.1103/PhysRevD.106.072012}}.

\bibitem{Chen:2008qw}
H.-X. Chen, A.~Hosaka, S.-L. Zhu, {$I^G J^{PC} = 1^- 1^{-+}$ tetraquark
  states}, Phys. Rev. D 78 (2008) 054017.
\newblock \href {http://arxiv.org/abs/0806.1998} {\path{arXiv:0806.1998}},
  \href {http://dx.doi.org/10.1103/PhysRevD.78.054017}
  {\path{doi:10.1103/PhysRevD.78.054017}}.

\bibitem{Chen:2008ne}
H.-X. Chen, A.~Hosaka, S.-L. Zhu, {$I^G J^{PC} = 0^+ 1^{-+}$ tetraquark
  states}, Phys. Rev. D 78 (2008) 117502.
\newblock \href {http://arxiv.org/abs/0808.2344} {\path{arXiv:0808.2344}},
  \href {http://dx.doi.org/10.1103/PhysRevD.78.117502}
  {\path{doi:10.1103/PhysRevD.78.117502}}.

\bibitem{Zhang:2019ykd}
X.~Zhang, J.-J. Xie, {Prediction of possible exotic states in the $\eta
  \bar{K}K^*$ system}, Chin. Phys. C 44~(5) (2020) 054104.
\newblock \href {http://arxiv.org/abs/1906.07340} {\path{arXiv:1906.07340}},
  \href {http://dx.doi.org/10.1088/1674-1137/44/5/054104}
  {\path{doi:10.1088/1674-1137/44/5/054104}}.

\bibitem{Dong:2022cuw}
X.-K. Dong, Y.-H. Lin, B.-S. Zou, {Interpretation of the $\eta_1(1855)$ as a
  $K\bar K_1(1400)+$ c.c. molecule}, Sci. China Phys. Mech. Astron. 65~(6)
  (2022) 261011.
\newblock \href {http://arxiv.org/abs/2202.00863} {\path{arXiv:2202.00863}},
  \href {http://dx.doi.org/10.1007/s11433-022-1887-5}
  {\path{doi:10.1007/s11433-022-1887-5}}.

\bibitem{Yang:2022lwq}
F.~Yang, H.~Q. Zhu, Y.~Huang, {Analysis of the $\eta_1(1855)$ as a
  $K\bar{K}_1(1400)$ molecular state}, Nucl. Phys. A 1030 (2023) 122571.
\newblock \href {http://arxiv.org/abs/2203.06934} {\path{arXiv:2203.06934}},
  \href {http://dx.doi.org/10.1016/j.nuclphysa.2022.122571}
  {\path{doi:10.1016/j.nuclphysa.2022.122571}}.

\bibitem{Wan:2022xkx}
B.-D. Wan, S.-Q. Zhang, C.-F. Qiao, {Possible structure of the newly found
  exotic state $\eta_1(1855)$}, Phys. Rev. D 106~(7) (2022) 074003.
\newblock \href {http://arxiv.org/abs/2203.14014} {\path{arXiv:2203.14014}},
  \href {http://dx.doi.org/10.1103/PhysRevD.106.074003}
  {\path{doi:10.1103/PhysRevD.106.074003}}.

\bibitem{Wang:2022sib}
X.-Y. Wang, F.-C. Zeng, X.~Liu, {Production of the $\eta_1(1855)$ through kaon
  induced reactions under the assumptions that it is a molecular or a hybrid
  state}, Phys. Rev. D 106~(3) (2022) 036005.
\newblock \href {http://arxiv.org/abs/2205.09283} {\path{arXiv:2205.09283}},
  \href {http://dx.doi.org/10.1103/PhysRevD.106.036005}
  {\path{doi:10.1103/PhysRevD.106.036005}}.

\bibitem{Su:2022eun}
N.~Su, H.-X. Chen, {$S$- and $P$-wave fully strange tetraquark states from QCD
  sum rules}, Phys. Rev. D 106~(1) (2022) 014023.
\newblock \href {http://arxiv.org/abs/2204.13959} {\path{arXiv:2204.13959}},
  \href {http://dx.doi.org/10.1103/PhysRevD.106.014023}
  {\path{doi:10.1103/PhysRevD.106.014023}}.

\bibitem{Yu:2022wtu}
Y.~Yu, X.~Zhuang, B.-C. Ke, Y.~Teng, Q.-S. Liu, {Investigating
  $\eta^\prime_{1}(1855)$ exotic states in
  $J/\psi\to\eta^\prime_{1}(1855)\eta^{(\prime)}$ decays}\href
  {http://arxiv.org/abs/2208.05442} {\path{arXiv:2208.05442}}.

\bibitem{Barnes:1977hg}
T.~Barnes, {Coloured quark and gluon constituents in the MIT bag model: A model
  of mesons}, Nucl. Phys. B 158 (1979) 171--188.
\newblock \href {http://dx.doi.org/10.1016/0550-3213(79)90194-9}
  {\path{doi:10.1016/0550-3213(79)90194-9}}.

\bibitem{Hasenfratz:1980jv}
P.~Hasenfratz, R.~R. Horgan, J.~Kuti, J.~M. Richard, {The effects of coloured
  glue in the QCD motivated bag of heavy quark-antiquark systems}, Phys. Lett.
  B 95 (1980) 299--305.
\newblock \href {http://dx.doi.org/10.1016/0370-2693(80)90491-8}
  {\path{doi:10.1016/0370-2693(80)90491-8}}.

\bibitem{Chanowitz:1982qj}
M.~S. Chanowitz, S.~R. Sharpe, {Hybrids: Mixed states of quarks and gluons},
  Nucl. Phys. B 222 (1983) 211--244, [Erratum: Nucl.Phys.B 228, 588--588
  (1983)].
\newblock \href {http://dx.doi.org/10.1016/0550-3213(83)90635-1}
  {\path{doi:10.1016/0550-3213(83)90635-1}}.

\bibitem{Isgur:1983wj}
N.~Isgur, J.~E. Paton, {A flux tube model for hadrons}, Phys. Lett. B 124
  (1983) 247--251.
\newblock \href {http://dx.doi.org/10.1016/0370-2693(83)91445-4}
  {\path{doi:10.1016/0370-2693(83)91445-4}}.

\bibitem{Close:1994hc}
F.~E. Close, P.~R. Page, {The production and decay of hybrid mesons by
  flux-tube breaking}, Nucl. Phys. B 443 (1995) 233--254.
\newblock \href {http://arxiv.org/abs/hep-ph/9411301}
  {\path{arXiv:hep-ph/9411301}}, \href
  {http://dx.doi.org/10.1016/0550-3213(95)00085-7}
  {\path{doi:10.1016/0550-3213(95)00085-7}}.

\bibitem{Page:1998gz}
P.~R. Page, E.~S. Swanson, A.~P. Szczepaniak, {Hybrid meson decay
  phenomenology}, Phys. Rev. D 59 (1999) 034016.
\newblock \href {http://arxiv.org/abs/hep-ph/9808346}
  {\path{arXiv:hep-ph/9808346}}, \href
  {http://dx.doi.org/10.1103/PhysRevD.59.034016}
  {\path{doi:10.1103/PhysRevD.59.034016}}.

\bibitem{Burns:2006wz}
T.~Burns, F.~E. Close, {Hybrid-meson properties in lattice QCD and flux-tube
  models}, Phys. Rev. D 74 (2006) 034003.
\newblock \href {http://arxiv.org/abs/hep-ph/0604161}
  {\path{arXiv:hep-ph/0604161}}, \href
  {http://dx.doi.org/10.1103/PhysRevD.74.034003}
  {\path{doi:10.1103/PhysRevD.74.034003}}.

\bibitem{Qiu:2022ktc}
L.~Qiu, Q.~Zhao, {Towards the establishment of the light $J^{P(C)}=1^{-(+)}$
  hybrid nonet}, Chin. Phys. C 46~(8) (2022) 051001.
\newblock \href {http://arxiv.org/abs/2202.00904} {\path{arXiv:2202.00904}},
  \href {http://dx.doi.org/10.1088/1674-1137/ac567e}
  {\path{doi:10.1088/1674-1137/ac567e}}.

\bibitem{Szczepaniak:2001rg}
A.~P. Szczepaniak, E.~S. Swanson, {Coulomb gauge QCD, confinement, and the
  constituent representation}, Phys. Rev. D 65 (2001) 025012.
\newblock \href {http://arxiv.org/abs/hep-ph/0107078}
  {\path{arXiv:hep-ph/0107078}}, \href
  {http://dx.doi.org/10.1103/PhysRevD.65.025012}
  {\path{doi:10.1103/PhysRevD.65.025012}}.

\bibitem{Iddir:2007dq}
F.~Iddir, L.~Semlala, {Hybrid states from constituent glue model}, Int. J. Mod.
  Phys. A 23 (2008) 5229--5250.
\newblock \href {http://arxiv.org/abs/0710.5352} {\path{arXiv:0710.5352}},
  \href {http://dx.doi.org/10.1142/S0217751X08042687}
  {\path{doi:10.1142/S0217751X08042687}}.

\bibitem{Guo:2007sm}
P.~Guo, A.~P. Szczepaniak, G.~Galata, A.~Vassallo, E.~Santopinto, {Gluelump
  spectrum from Coulomb gauge QCD}, Phys. Rev. D 77 (2008) 056005.
\newblock \href {http://arxiv.org/abs/0707.3156} {\path{arXiv:0707.3156}},
  \href {http://dx.doi.org/10.1103/PhysRevD.77.056005}
  {\path{doi:10.1103/PhysRevD.77.056005}}.

\bibitem{Andreev:2012hw}
O.~Andreev, {Exotic hybrid pseudopotentials and gauge/string duality}, Phys.
  Rev. D 87~(6) (2013) 065006.
\newblock \href {http://arxiv.org/abs/1211.0930} {\path{arXiv:1211.0930}},
  \href {http://dx.doi.org/10.1103/PhysRevD.87.065006}
  {\path{doi:10.1103/PhysRevD.87.065006}}.

\bibitem{Bellantuono:2014lra}
L.~Bellantuono, P.~Colangelo, F.~Giannuzzi, {Exotic $J^{PC}=1^{-+}$ mesons in a
  holographic model of QCD}, Eur. Phys. J. C 74~(4) (2014) 2830.
\newblock \href {http://arxiv.org/abs/1402.5308} {\path{arXiv:1402.5308}},
  \href {http://dx.doi.org/10.1140/epjc/s10052-014-2830-6}
  {\path{doi:10.1140/epjc/s10052-014-2830-6}}.

\bibitem{Xu:2018cor}
S.-S. Xu, Z.-F. Cui, L.~Chang, J.~Papavassiliou, C.~D. Roberts, H.-S. Zong,
  {New perspective on hybrid mesons}, Eur. Phys. J. A 55~(7) (2019) 113.
\newblock \href {http://arxiv.org/abs/1805.06430} {\path{arXiv:1805.06430}},
  \href {http://dx.doi.org/10.1140/epja/i2019-12805-4}
  {\path{doi:10.1140/epja/i2019-12805-4}}.

\bibitem{Michael:1985ne}
C.~Michael, {Adjoint sources in lattice gauge theory}, Nucl. Phys. B 259 (1985)
  58--76.
\newblock \href {http://dx.doi.org/10.1016/0550-3213(85)90297-4}
  {\path{doi:10.1016/0550-3213(85)90297-4}}.

\bibitem{McNeile:1998cp}
C.~McNeile, C.~W. Bernard, T.~A. DeGrand, C.~E. DeTar, S.~A. Gottlieb, U.~M.
  Heller, J.~Hetrick, R.~Sugar, D.~Toussaint, {Exotic meson spectroscopy from
  the clover action at $\beta = 5.85$ and $\beta = 6.15$}, Nucl. Phys. B Proc.
  Suppl. 73 (1999) 264--266.
\newblock \href {http://arxiv.org/abs/hep-lat/9809087}
  {\path{arXiv:hep-lat/9809087}}, \href
  {http://dx.doi.org/10.1016/S0920-5632(99)85043-9}
  {\path{doi:10.1016/S0920-5632(99)85043-9}}.

\bibitem{Juge:2002br}
K.~J. Juge, J.~Kuti, C.~Morningstar, {Fine Structure of the QCD String
  Spectrum}, Phys. Rev. Lett. 90 (2003) 161601.
\newblock \href {http://arxiv.org/abs/hep-lat/0207004}
  {\path{arXiv:hep-lat/0207004}}, \href
  {http://dx.doi.org/10.1103/PhysRevLett.90.161601}
  {\path{doi:10.1103/PhysRevLett.90.161601}}.

\bibitem{Lacock:1996ny}
P.~Lacock, C.~Michael, P.~Boyle, P.~Rowland, {Hybrid mesons from quenched QCD},
  Phys. Lett. B 401 (1997) 308--312.
\newblock \href {http://arxiv.org/abs/hep-lat/9611011}
  {\path{arXiv:hep-lat/9611011}}, \href
  {http://dx.doi.org/10.1016/S0370-2693(97)00384-5}
  {\path{doi:10.1016/S0370-2693(97)00384-5}}.

\bibitem{MILC:1997usn}
C.~W. Bernard, et~al., {Exotic mesons in quenched lattice QCD}, Phys. Rev. D 56
  (1997) 7039--7051.
\newblock \href {http://arxiv.org/abs/hep-lat/9707008}
  {\path{arXiv:hep-lat/9707008}}, \href
  {http://dx.doi.org/10.1103/PhysRevD.56.7039}
  {\path{doi:10.1103/PhysRevD.56.7039}}.

\bibitem{Bernard:2003jd}
C.~Bernard, T.~Burch, E.~B. Gregory, D.~Toussaint, C.~E. DeTar, J.~Osborn,
  S.~A. Gottlieb, U.~M. Heller, R.~Sugar, {Lattice calculation of $1^{-+}$
  hybrid mesons with improved Kogut-Susskind fermions}, Phys. Rev. D 68 (2003)
  074505.
\newblock \href {http://arxiv.org/abs/hep-lat/0301024}
  {\path{arXiv:hep-lat/0301024}}, \href
  {http://dx.doi.org/10.1103/PhysRevD.68.074505}
  {\path{doi:10.1103/PhysRevD.68.074505}}.

\bibitem{Hedditch:2005zf}
J.~N. Hedditch, W.~Kamleh, B.~G. Lasscock, D.~B. Leinweber, A.~G. Williams,
  J.~M. Zanotti, {$1^{-+}$ exotic meson at light quark masses}, Phys. Rev. D 72
  (2005) 114507.
\newblock \href {http://arxiv.org/abs/hep-lat/0509106}
  {\path{arXiv:hep-lat/0509106}}, \href
  {http://dx.doi.org/10.1103/PhysRevD.72.114507}
  {\path{doi:10.1103/PhysRevD.72.114507}}.

\bibitem{Dudek:2009qf}
J.~J. Dudek, R.~G. Edwards, M.~J. Peardon, D.~G. Richards, C.~E. Thomas,
  {Highly Excited and Exotic Meson Spectrum from Dynamical Lattice QCD}, Phys.
  Rev. Lett. 103 (2009) 262001.
\newblock \href {http://arxiv.org/abs/0909.0200} {\path{arXiv:0909.0200}},
  \href {http://dx.doi.org/10.1103/PhysRevLett.103.262001}
  {\path{doi:10.1103/PhysRevLett.103.262001}}.

\bibitem{Dudek:2010wm}
J.~J. Dudek, R.~G. Edwards, M.~J. Peardon, D.~G. Richards, C.~E. Thomas,
  {Toward the excited meson spectrum of dynamical QCD}, Phys. Rev. D 82 (2010)
  034508.
\newblock \href {http://arxiv.org/abs/1004.4930} {\path{arXiv:1004.4930}},
  \href {http://dx.doi.org/10.1103/PhysRevD.82.034508}
  {\path{doi:10.1103/PhysRevD.82.034508}}.

\bibitem{Dudek:2013yja}
J.~J. Dudek, R.~G. Edwards, P.~Guo, C.~E. Thomas, {Toward the excited isoscalar
  meson spectrum from lattice QCD}, Phys. Rev. D 88~(9) (2013) 094505.
\newblock \href {http://arxiv.org/abs/1309.2608} {\path{arXiv:1309.2608}},
  \href {http://dx.doi.org/10.1103/PhysRevD.88.094505}
  {\path{doi:10.1103/PhysRevD.88.094505}}.

\bibitem{Chen:2022isv}
F.~Chen, X.~Jiang, Y.~Chen, M.~Gong, Z.~Liu, C.~Shi, W.~Sun, {$1^{-+}$ Hybrid
  in $J/\psi$ Radiative Decays from Lattice QCD}\href
  {http://arxiv.org/abs/2207.04694} {\path{arXiv:2207.04694}}.

\bibitem{Balitsky:1982ps}
I.~I. Balitsky, D.~Diakonov, A.~V. Yung, {Exotic mesons with $J^{PC} = 1^{-+}$
  from QCD sum rules}, Phys. Lett. B 112 (1982) 71--75.
\newblock \href {http://dx.doi.org/10.1016/0370-2693(82)90908-X}
  {\path{doi:10.1016/0370-2693(82)90908-X}}.

\bibitem{Govaerts:1983ka}
J.~Govaerts, F.~de~Viron, D.~Gusbin, J.~Weyers, {Exotic mesons from QCD sum
  rules}, Phys. Lett. B 128 (1983) 262, [Erratum: Phys.Lett.B 136, 445--445
  (1984)].
\newblock \href {http://dx.doi.org/10.1016/0370-2693(83)90405-7}
  {\path{doi:10.1016/0370-2693(83)90405-7}}.

\bibitem{Kisslinger:1995yw}
L.~S. Kisslinger, Z.~P. Li, {Hybrid baryons via QCD sum rules}, Phys. Rev. D 51
  (1995) R5986--R5989.
\newblock \href {http://dx.doi.org/10.1103/PhysRevD.51.R5986}
  {\path{doi:10.1103/PhysRevD.51.R5986}}.

\bibitem{Chetyrkin:2000tj}
K.~G. Chetyrkin, S.~Narison, {Light hybrid mesons in QCD}, Phys. Lett. B 485
  (2000) 145--150.
\newblock \href {http://arxiv.org/abs/hep-ph/0003151}
  {\path{arXiv:hep-ph/0003151}}, \href
  {http://dx.doi.org/10.1016/S0370-2693(00)00621-3}
  {\path{doi:10.1016/S0370-2693(00)00621-3}}.

\bibitem{Jin:2002rw}
H.~Y. Jin, J.~G. Korner, T.~G. Steele, {Improved determination of the mass of
  the $1^{-+}$ light hybrid meson from QCD sum rules}, Phys. Rev. D 67 (2003)
  014025.
\newblock \href {http://arxiv.org/abs/hep-ph/0211304}
  {\path{arXiv:hep-ph/0211304}}, \href
  {http://dx.doi.org/10.1103/PhysRevD.67.014025}
  {\path{doi:10.1103/PhysRevD.67.014025}}.

\bibitem{Narison:2009vj}
S.~Narison, {$1^{-+}$ light exotic mesons in QCD}, Phys. Lett. B 675 (2009)
  319--325.
\newblock \href {http://arxiv.org/abs/0903.2266} {\path{arXiv:0903.2266}},
  \href {http://dx.doi.org/10.1016/j.physletb.2009.04.012}
  {\path{doi:10.1016/j.physletb.2009.04.012}}.

\bibitem{Huang:2010dc}
P.-Z. Huang, H.-X. Chen, S.-L. Zhu, {Strong decay patterns of the $1^{-+}$
  exotic hybrid mesons}, Phys. Rev. D 83 (2011) 014021.
\newblock \href {http://arxiv.org/abs/1010.2293} {\path{arXiv:1010.2293}},
  \href {http://dx.doi.org/10.1103/PhysRevD.83.014021}
  {\path{doi:10.1103/PhysRevD.83.014021}}.

\bibitem{Chen:2010ic}
H.-X. Chen, Z.-X. Cai, P.-Z. Huang, S.-L. Zhu, {Decay properties of the
  $1^{-+}$ hybrid state}, Phys. Rev. D 83 (2011) 014006.
\newblock \href {http://arxiv.org/abs/1010.3974} {\path{arXiv:1010.3974}},
  \href {http://dx.doi.org/10.1103/PhysRevD.83.014006}
  {\path{doi:10.1103/PhysRevD.83.014006}}.

\bibitem{Huang:2016upt}
Z.-R. Huang, H.-Y. Jin, T.~G. Steele, Z.-F. Zhang, {Revisiting the $b_1\pi$ and
  $\rho\pi$ decay modes of the $1^{-+}$ light hybrid state with light-cone QCD
  sum rules}, Phys. Rev. D 94~(5) (2016) 054037.
\newblock \href {http://arxiv.org/abs/1608.03028} {\path{arXiv:1608.03028}},
  \href {http://dx.doi.org/10.1103/PhysRevD.94.054037}
  {\path{doi:10.1103/PhysRevD.94.054037}}.

\bibitem{Li:2021fwk}
S.-H. Li, Z.-S. Chen, H.-Y. Jin, W.~Chen, {Mass of $1^{-+}$ fourquark-hybrid
  mixed states}, Phys. Rev. D 105~(5) (2022) 054030.
\newblock \href {http://arxiv.org/abs/2111.13897} {\path{arXiv:2111.13897}},
  \href {http://dx.doi.org/10.1103/PhysRevD.105.054030}
  {\path{doi:10.1103/PhysRevD.105.054030}}.

\bibitem{Shastry:2022mhk}
V.~Shastry, C.~S. Fischer, F.~Giacosa, {The phenomenology of the exotic hybrid
  nonet with $\pi_1(1600)$ and $\eta_1(1855)$}, Phys. Lett. B 834 (2022)
  137478.
\newblock \href {http://arxiv.org/abs/2203.04327} {\path{arXiv:2203.04327}},
  \href {http://dx.doi.org/10.1016/j.physletb.2022.137478}
  {\path{doi:10.1016/j.physletb.2022.137478}}.

\bibitem{Horn:1977rq}
D.~Horn, J.~Mandula, {Model of mesons with constituent gluons}, Phys. Rev. D 17
  (1978) 898.
\newblock \href {http://dx.doi.org/10.1103/PhysRevD.17.898}
  {\path{doi:10.1103/PhysRevD.17.898}}.

\bibitem{Coyne:1980zd}
J.~J. Coyne, P.~M. Fishbane, S.~Meshkov, {Glueballs: Their spectra, production
  and decay}, Phys. Lett. B 91 (1980) 259.
\newblock \href {http://dx.doi.org/10.1016/0370-2693(80)90445-1}
  {\path{doi:10.1016/0370-2693(80)90445-1}}.

\bibitem{Chanowitz:1980gu}
M.~S. Chanowitz, {Have We Seen Our First Glueball?}, Phys. Rev. Lett. 46 (1981)
  981.
\newblock \href {http://dx.doi.org/10.1103/PhysRevLett.46.981}
  {\path{doi:10.1103/PhysRevLett.46.981}}.

\bibitem{Barnes:1981ac}
T.~Barnes, {A transverse gluonium potential model with Breit-fermi hyperfine
  effects}, Z. Phys. C 10 (1981) 275.
\newblock \href {http://dx.doi.org/10.1007/BF01549736}
  {\path{doi:10.1007/BF01549736}}.

\bibitem{Cornwall:1982zn}
J.~M. Cornwall, A.~Soni, {Glueballs as bound states of massive gluons}, Phys.
  Lett. B 120 (1983) 431.
\newblock \href {http://dx.doi.org/10.1016/0370-2693(83)90481-1}
  {\path{doi:10.1016/0370-2693(83)90481-1}}.

\bibitem{Cho:2015rsa}
Y.~M. Cho, X.~Y. Pham, P.~Zhang, J.-J. Xie, L.-P. Zou, {Glueball physics in
  QCD}, Phys. Rev. D 91~(11) (2015) 114020.
\newblock \href {http://arxiv.org/abs/1503.08890} {\path{arXiv:1503.08890}},
  \href {http://dx.doi.org/10.1103/PhysRevD.91.114020}
  {\path{doi:10.1103/PhysRevD.91.114020}}.

\bibitem{Chen:2021smz}
H.-X. Chen, W.~Chen, S.-L. Zhu, {New hadron configuration: The double-gluon
  hybrid state}, Phys. Rev. D 105~(5) (2022) L051501.
\newblock \href {http://arxiv.org/abs/2111.04514} {\path{arXiv:2111.04514}},
  \href {http://dx.doi.org/10.1103/PhysRevD.105.L051501}
  {\path{doi:10.1103/PhysRevD.105.L051501}}.

\bibitem{Tang:2021zti}
C.-M. Tang, Y.-C. Zhao, L.~Tang, {Mass predictions of vector ($1^{--}$)
  double-gluon heavy quarkonium hybrids from QCD sum rules}, Phys. Rev. D
  105~(11) (2022) 114004.
\newblock \href {http://arxiv.org/abs/2111.07328} {\path{arXiv:2111.07328}},
  \href {http://dx.doi.org/10.1103/PhysRevD.105.114004}
  {\path{doi:10.1103/PhysRevD.105.114004}}.

\bibitem{Ovchinnikov:1988gk}
A.~A. Ovchinnikov, A.~A. Pivovarov, {QCD sum rule calculation of the quark
  gluon condensate}, Sov. J. Nucl. Phys. 48 (1988) 721--723.

\bibitem{Yang:1993bp}
K.-C. Yang, W.~Y.~P. Hwang, E.~M. Henley, L.~S. Kisslinger, {QCD sum rules and
  neutron-proton mass difference}, Phys. Rev. D 47 (1993) 3001--3012.
\newblock \href {http://dx.doi.org/10.1103/PhysRevD.47.3001}
  {\path{doi:10.1103/PhysRevD.47.3001}}.

\bibitem{Ellis:1996xc}
J.~R. Ellis, E.~Gardi, M.~Karliner, M.~A. Samuel, {Renormalization-scheme
  dependence of Pad{\'e} summation in QCD}, Phys. Rev. D 54 (1996) 6986--6996.
\newblock \href {http://arxiv.org/abs/hep-ph/9607404}
  {\path{arXiv:hep-ph/9607404}}, \href
  {http://dx.doi.org/10.1103/PhysRevD.54.6986}
  {\path{doi:10.1103/PhysRevD.54.6986}}.

\bibitem{Ioffe:2002be}
B.~L. Ioffe, K.~N. Zyablyuk, {Gluon condensate in charmonium sum rules with
  three-loop corrections}, Eur. Phys. J. C 27 (2003) 229--241.
\newblock \href {http://arxiv.org/abs/hep-ph/0207183}
  {\path{arXiv:hep-ph/0207183}}, \href
  {http://dx.doi.org/10.1140/epjc/s2002-01099-8}
  {\path{doi:10.1140/epjc/s2002-01099-8}}.

\bibitem{Jamin:2002ev}
M.~Jamin, {Flavor-symmetry breaking of the quark condensate and chiral
  corrections to the Gell-Mann-Oakes-Renner relation}, Phys. Lett. B 538 (2002)
  71--76.
\newblock \href {http://arxiv.org/abs/hep-ph/0201174}
  {\path{arXiv:hep-ph/0201174}}, \href
  {http://dx.doi.org/10.1016/S0370-2693(02)01951-2}
  {\path{doi:10.1016/S0370-2693(02)01951-2}}.

\bibitem{Gimenez:2005nt}
V.~Gimenez, V.~Lubicz, F.~Mescia, V.~Porretti, J.~Reyes, {Operator product
  expansion and quark condensate from lattice QCD in coordinate space}, Eur.
  Phys. J. C 41 (2005) 535--544.
\newblock \href {http://arxiv.org/abs/hep-lat/0503001}
  {\path{arXiv:hep-lat/0503001}}, \href
  {http://dx.doi.org/10.1140/epjc/s2005-02250-9}
  {\path{doi:10.1140/epjc/s2005-02250-9}}.

\bibitem{Narison:2011xe}
S.~Narison, {Gluon condensates and precise $\overline{m}_{c,b}$ from
  QCD-moments and their ratios to order $\alpha_s^3$ and $\langle G^4
  \rangle$}, Phys. Lett. B 706 (2012) 412--422.
\newblock \href {http://arxiv.org/abs/1105.2922} {\path{arXiv:1105.2922}},
  \href {http://dx.doi.org/10.1016/j.physletb.2011.11.058}
  {\path{doi:10.1016/j.physletb.2011.11.058}}.

\bibitem{Narison:2018dcr}
S.~Narison, {QCD parameter correlations from heavy quarkonia}, Int. J. Mod.
  Phys. A 33~(10) (2018) 1850045, [Addendum: Int.J.Mod.Phys.A 33, 1892004
  (2018)].
\newblock \href {http://arxiv.org/abs/1801.00592} {\path{arXiv:1801.00592}},
  \href {http://dx.doi.org/10.1142/S0217751X18500458}
  {\path{doi:10.1142/S0217751X18500458}}.

\end{thebibliography}

\end{document}